\documentclass[fleqn,usenatbib]{mnras}

\usepackage[T1]{fontenc}
\usepackage{ae,aecompl}

\usepackage{graphicx}	
\usepackage{amsmath}	
\usepackage{amssymb}	
\usepackage[percent]{overpic}

\title[Retrieval of Exoplanet Emission Spectra with HyDRA]{Retrieval of Exoplanet Emission Spectra with HyDRA}
\author[Gandhi \& Madhusudhan]{
Siddharth Gandhi$^{1}$\thanks{E-mail: sng29@ast.cam.ac.uk} \&
Nikku Madhusudhan$^{1}$\thanks{E-mail: nmadhu@ast.cam.ac.uk}
\\
$^{1}$Institute of Astronomy, University of Cambridge, Madingley Road, Cambridge, CB3 0HA, UK
}

\date{Accepted XXX. Received YYY; in original form ZZZ}

\pubyear{2017}

\begin{document}
\label{firstpage}
\pagerange{\pageref{firstpage}--\pageref{lastpage}}
\maketitle

\begin{abstract}
Thermal emission spectra of exoplanets provide constraints on the chemical compositions, pressure-temperature ($P$-$T$) profiles, and energy transport in exoplanetary atmospheres. Accurate inferences of these properties rely on the robustness of the atmospheric retrieval methods employed. While extant retrieval codes have provided significant constraints on molecular abundances and temperature profiles in several exoplanetary atmospheres, the constraints on their deviations from thermal and chemical equilibria have yet to be fully explored. Our present work is a step in this direction. We report HyDRA, a disequilibrium retrieval framework for thermal emission spectra of exoplanetary atmospheres. The retrieval code uses the standard architecture of a parametric atmospheric model coupled with Bayesian statistical inference using the Nested Sampling algorithm. For a given dataset, the retrieved compositions and $P$-$T$ profiles are used in tandem with the GENESIS self-consistent atmospheric model to constrain layer-by-layer deviations from chemical and radiative-convective equilibrium in the observable atmosphere. We demonstrate HyDRA on the Hot Jupiter WASP-43b with a high-precision emission spectrum. We retrieve an H$_2$O mixing ratio of $\log{\rm (H_2O)} = -3.54^{+0.82}_{-0.52}$, consistent with previous studies. We detect H$_2$O and a combined CO/CO$_2$ at 8-$\sigma$ significance. We find the dayside $P$-$T$ profile to be consistent with radiative-convective equilibrium within the 1-$\sigma$ limits and with low day-night redistribution, consistent with previous studies. The derived compositions are also consistent with thermochemical equilibrium for the corresponding distribution of $P$-$T$ profiles. In the era of high precision and high resolution emission spectroscopy, HyDRA provides a path to retrieve disequilibrium phenomena in exoplanetary atmospheres. 
\end{abstract}




\section{Introduction}
\label{introduction}

Atmospheric studies are at the forefront of exoplanetary research. Recent observational efforts have led to significant advances in our understanding of exoplanetary atmospheres \citep{madhu_2016}. The transit method has been the most successful to date with spectra of numerous planets observed \citep{deming_2013,kreidberg_2014}, but high quality spectra have also been observed using direct imaging \citep{konopacky_2013,macintosh_2015} and very high resolution spectroscopy \citep{snellen_2010,birkby_2013}. Hot Jupiter atmospheres in particular are prime candidates for atmospheric characterisation, thanks to their extended atmospheres and strong thermal emission, resulting in strong spectral signatures. Spectra of transiting hot giant planets have thus led to significant constraints on the chemical compositions, temperature profiles, clouds/hazes, and other properties of their atmospheres \citep{madhu_2011_inv,kreidberg_2014,stevenson_2014,sing_2016}. The compositional estimates are in turn beginning to provide constraints not only on the atmospheric processes but also on their formation processes \citep{madhu_2014c,lavie_2017}. 

The transit method, in particular, allows probing of the planetary atmosphere in multiple configurations. A transmission spectrum, obtained when the planet transits in front of the star, probes the atmosphere at the day-night terminator region of the planet. On the other hand, an emission spectrum obtained during secondary eclipse when the planet passes behind the star probes the dayside atmosphere of the planet. While constraints on various atmospheric properties have been placed using both configurations, emission spectra are particularly conducive to place detailed constraints on the composition as well as the temperature structure of the dayside atmosphere. High precision near-infrared emission spectra of transiting exoplanets observed with HST, Spitzer, and ground-based facilities have indeed provided new insights into their temperature structures. The imminent arrival of JWST is expected to further enhance our ability to constrain such physical and chemical properties.

Atmospheric retrieval involves deriving the atmospheric properties of an exoplanet given an observed spectrum. This involves parameter estimation of a model atmosphere from a spectral dataset using detailed statistical inference methods. Several studies have conducted atmospheric retrieval of thermal emission spectra of transiting exoplanets \citep{madhu_2009,madhu_2011_inv,lee_2012, line_2012,evans_2017,oreshenko_2017}. The methods generally involve a parametric model, with the pressure-temperature ($P$-$T$) profile and abundances of chemical species as free parameters, with no prior assumptions about chemical or radiative equilibrium. The model is coupled to a statistical inference algorithm to explore the model space and estimate the parameters. A variety of statistical methods have been used in the literature with varying levels of sophistication and have resulted in constraints on thermal inversions and chemical abundances in several exoplanets \citep{madhu_2011_inv,line_2014,haynes_2015,evans_2017}. 

It is important to distinguish the parametric models used in retrieval to self-consistent equilibrium models. The latter models compute the $P$-$T$ profile, molecular composition, and the spectrum, of an atmosphere ab initio based on assumptions of thermochemical and radiative-convective equilibrium given the macroscopic system parameters \cite[see e.g.][for a comparison between retrieval and equilibrium models]{madhu_2014}. Such models are valuable to simulate and investigate physical processes in exoplanetary atmospheres, to predict observables  and to define the limits of our theoretical understanding. The complexity of these models varies significantly, ranging from 1-D atmospheres \citep[e.g.][]{seager_2005,burrows_2008,fortney_2008,molliere_2015,malik_2017,gandhi_2017} to full 3-D general circulation models \citep[e.g.][]{showman_2009,kataria_2015}. While such forward models are highly beneficial they are limited in their capability to directly interpret observations. Therefore, both self-consistent models and retrieval methods are important for detailed characterisation of exoplanetary atmospheres. 

What is ultimately desirable is a self-consistent equilibrium model that can work in tandem with a retrieval. The primary advantage of retrieval methods is the ability to estimate the composition and the P-T profiles from spectral data without any a priori assumptions, e.g of  chemical/radiative equilibrium. Conversely, retrievals can in principle also be used to constrain deviations of the retrieved thermal/chemical properties from equilibrium expectations, thereby allowing constraints on non-equilibrium processes. Some studies in the past have explored this avenue to constrain disequilibrium chemistry in exoplanetary atmospheres. Early retrievals used a chemical parametrisation which directly retrieved deviations from equilibrium chemistry \citep{madhu_2009}. More recently, retrievals typically use parametric mixing ratios, assumed to be uniform in the atmosphere, and assess for deviation from chemical equilibrium a posteriori \citep{stevenson_2010,madhu_2011}. On the other hand, some studies have considered enforcing chemical equilibrium to narrow down the solution space in retrievals, referred to as ``chemically consistent" retrievals \citep{line_2016,kreidberg_2014}. However, statistical constraints on deviations from chemical equilibrium are not a routine feature in most retrievals. 

As for compositional disequilibrium, retrievals in principle should also be able to constrain deviations of the retrieved P-T profiles from radiative-convective equilibrium if any. To date this aspect has not been explored, arguably due to computational challenges in the past. Most notably, self-consistent versus retrieval modelling codes often employ different frameworks owing to their contrasting functionalities and their development by independent groups. Thusly compatibility between the models is often difficult. The computational time is also a consideration, as self-consistent models are typically significantly slower per model evaluation than parametric models used in retrievals. The current work aims to bring together both forward and retrieval methods into a common framework in order to facilitate simultaneous constraints on chemical and radiative disequilibrium. 

Here we introduce HyDRA, an integrated atmospheric retrieval framework for thermal emission spectroscopy of transiting exoplanets. In addition to retrieving chemical compositions and $P$-$T$ profiles, HyDRA allows constraints on layer-by-layer deviations of the retrieved atmosphere from chemical and radiative-convective equilibrium. This is pursued by integrating a custom-built retrieval code with the fully self-consistent equilibrium model GENESIS \citep{gandhi_2017}. Both the retrieval and equilibrium codes share the same structure, language, and underlying input data (e.g. opacity database, stellar flux, system parameters, etc.). This means that differences between our retrieved parameters and our equilibrium model will be down to the atmospheric processes at play, and not any differences between the modelling schemes. We study the disequilibrium for our test case of WASP-43b, with one of the most precise spectra available to date. We are able to show using HyDRA that WASP-43b is consistent with both chemical and radiative-convective equilibrium.

In what follows, we describe our modelling and retrieval methodology in  section~\ref{methods}. We validate the HyDRA retrieval framework using a synthetic data set in section~\ref{validation}. We then use HyDRA to retrieve the dayside atmospheric properties of the hot Jupiter WASP-43b in section \ref{results}, followed by a summary and discussion in section~\ref{discussion}. 

\section{Methods}
\label{methods}


\begin{figure*}
\includegraphics[width=\textwidth,height=1.2\columnwidth,trim=0.8cm 0 0 0,clip]{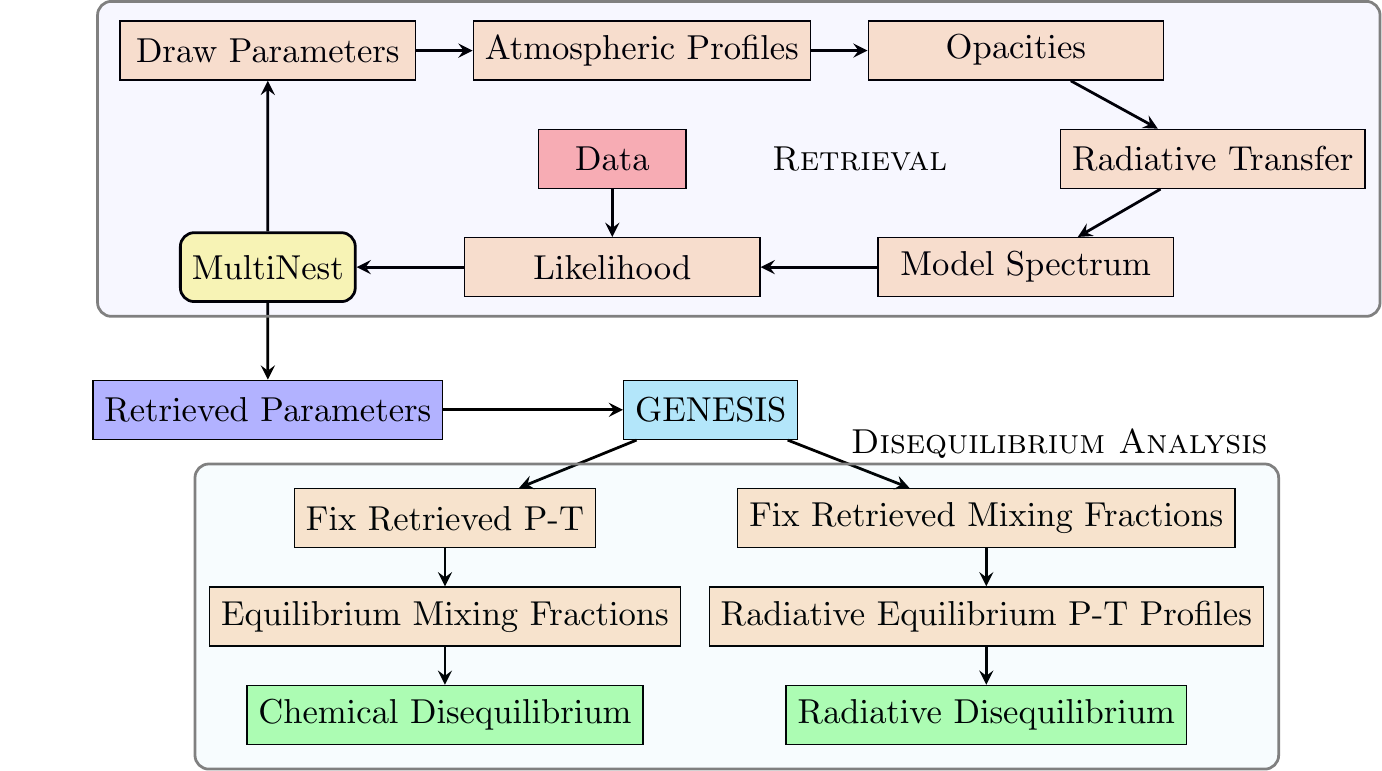}
\caption{The HyDRA modelling and retrieval framework. The model parameters, discussed in section \ref{methods_model_params}, are used to compute the atmospheric structure (see section \ref{methods_pt} and \ref{methods_mix_frac}), the opacities (section \ref{methods_opac}) and the emergent spectrum for a model atmosphere (section \ref{methods_rad_transfer} and \ref{methods_bin_data}). The likelihood is computed using the data and the model spectrum binned to the resolution of the data. The statistical inference, including parameter estimation and model selection, is conducted using the nested sampling algorithm implemented using the MultiNest package, as discussed in section \ref{multinest}. Once the retrieval is completed the retrieved P-T profile and chemical compositions are used in tandem with the GENESIS self-consistent equilibrium model to compute deviations from radiative-convective equilibrium and chemical equilibrium, as discussed in section \ref{methods_diseqm}.}
\label{fig:flowchart}
\end{figure*}

HyDRA is a custom-built atmospheric retrieval framework for application to emission spectra of exoplanets. The framework comprises of three key components: (1) a parametric atmosphere model, (2) a Bayesian statistical inference algorithm, and (3) a disequilibrium module for constraining deviations as seen from equilibrium. The parametric model computes an atmospheric thermal emission spectrum given the parametric composition and temperature structure. Given a dataset, the Bayesian inference involves estimating the model parameters and detection significances. The disequilibrium module constrains the deviations of the retrieved atmospheric properties from chemical and radiative-convective equilibrium. In the following sections, we discuss each of the above aspects of HyDRA. The modelling and retrieval architecture of HyDRA is shown in fig. \ref{fig:flowchart}.

\subsection{Geometry}
\label{geometry}

In the present work we focus on emission spectra of transiting exoplanets as observed at secondary eclipse. Immediately prior to secondary eclipse, emission is observed both from the star as well as the dayside of the planet combined. When the planet is in secondary eclipse only the stellar flux is observed, which when subtracted from the combined emission gives the planetary spectrum. Dividing the two quantities yields the planet-star flux ratio that is independent of the distance to the system. The observed planet-star flux ratio can be expressed as 
\begin{align}
\frac{\mathrm{F_\mathrm{p}(\nu)}}{\mathrm{F_{star}(\nu)}} & \approx \frac{\mathrm{R_p^2B(T_{p,\nu},\nu)}}{\mathrm{R_{star}^2B(T_{star,\nu},\nu)}}.
\end{align}
Here, T$_\mathrm{p,\nu}$ and T$_\mathrm{star,\nu}$ refer to the planetary and stellar brightness temperatures at the frequency $\nu$, and $R_\mathrm{p}$ and $R_\mathrm{star}$ are their corresponding radii. $B(T,\nu)$ is the Planck function corresponding to the brightness temperature $\mathrm{T,\nu}$ at a frequency $\nu$. T$_\mathrm{p,\nu}$ is a representative temperature corresponding to the $\tau_\nu \approx 1$ surface (the ``photosphere") at frequency $\nu$. The exact calculation of the emergent spectrum from the planetary atmosphere is described in section \ref{methods_rad_transfer}.

The flux ratio has wavelength dependant emission which occurs from different pressure levels in the atmosphere depending on the opacity and hence atmospheric chemistry. The emission spectrum therefore provides constraints on the temperature profile, chemical composition, and energy transport in the dayside atmosphere. Hot Jupiter atmospheres are particularly conducive to observations of thermal emission due to their large radii and high temperatures. 

\begin{figure}
\begin{overpic}[width=\columnwidth]{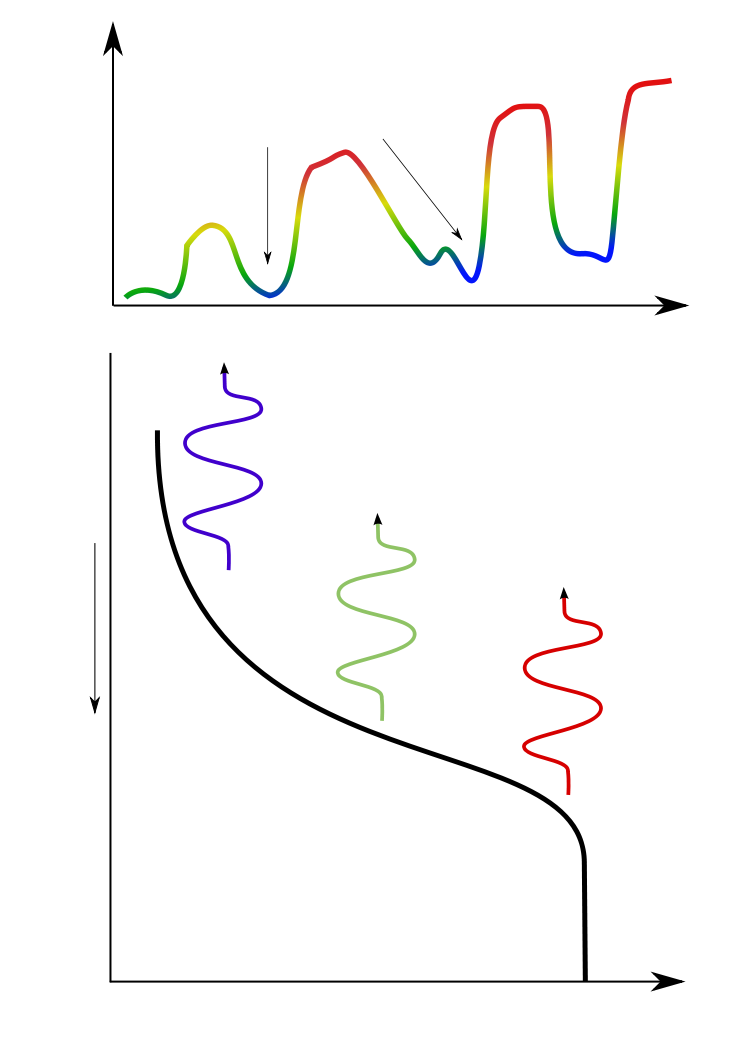}
 \put (60,2) {\LARGE $\mathrm{T}$}
 \put (60,67) {\LARGE $\lambda$}
 \put (19,20) {\large $\mathrm{Atmospheric \, \, P-T \, \, profile}$}
 \put (18,88) {\large $\mathrm{Absorption \, \, Features}$}
 \put (26,44) {\rotatebox{90}{\large $\mathrm{Low - T \, \, Emission}$}}
 \put (58,25) {\rotatebox{90}{\large $\mathrm{High - T \, \, Emission}$}}
 \put (4,80) {\rotatebox{90}{\LARGE $\mathrm{Flux}$}}
 \put (4,35) {\rotatebox{90}{\LARGE $\mathrm{log(P)}$}}
\end{overpic}
\caption{Schematic of thermal emission from an atmosphere. The lower diagram shows a model pressure-temperature profile and the upper diagram shows the corresponding observed spectrum. If the emission occurs from lower in the atmosphere (where the temperature is greater) the emitted flux is greater (red regions) and vice versa for cooler (blue) regions. Thus, features in molecular bands where the opacity is greater appear as absorption features in the flux spectrum if the temperature decreases with height and originate higher up in the atmosphere.}
\label{fig:sketch}
\end{figure}
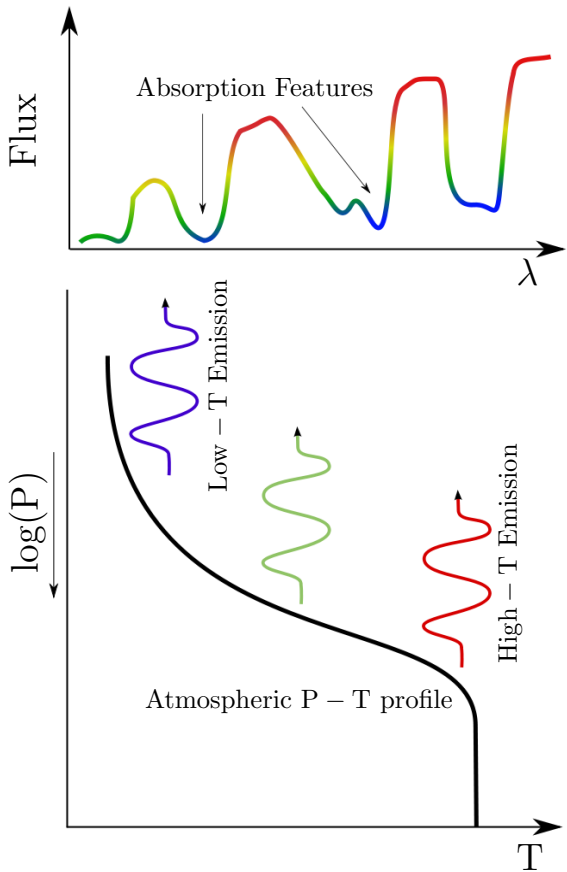

Figure \ref{fig:sketch} shows a schematic of the emission occurring from a planet, with the pressure-temperature (henceforth P-T) profile shown. The region of the atmosphere where the optical depth is of order unity is where the emission will occur from. This will also be a function of the wavelength and of the constituent molecular species (i.e their cross-section). Where the emission occurs from cooler parts of the atmosphere, a smaller thermal signal is generated. Absorption features thus result from the temperature decreasing with altitude, and vice versa for thermal inversions (where the temperature increases with height). This allows us to probe the P-T profile of the planet by studying the spectrum and the absorption from spectrally active species. In the retrieval algorithm, we will need to model the atmosphere for a wide range of P-T profiles and chemistry that is possible for such exoplanets to explore the parameter space.

In our model we take a grid of 4000 evenly spaced wavelength points in the Hubble WFC3 and Spitzer IRAC 1 and 2 bandpasses between 1 and 5.5$\micron$, and 100 atmospheric layers evenly spaced in log(P) ranging from $10^2-10^{-5}$ bar. The temperature profile is used to determine the density of gas under the ideal gas assumption, taking the atmosphere to be in hydrostatic equilibrium. 

\subsection{Model Parameters}
\label{methods_model_params}

\begin{table*}
	\centering
	\caption{The set of parameters we choose in order to test our model. These were chosen so as to explore the model in the region of the parameter space in which we would expect the results to lie and taken from a self-consistent equilibrium profile. The first 7 values are the mixing fractions of the molecular species (section \ref{methods_mix_frac}) and the total absorption coefficient calculation is given in section \ref{methods_opac}. The final six generate the P-T profile described in section \ref{methods_pt}. As the original P-T profile to generate the data set is from the self-consistent model, the analytic profile parameters are the closest fit.}
	\label{tab:params_table}
\begin{tabular}{llllll lllllllr}
\hline
\multicolumn{5}{c}{Mixing Fraction} &\multicolumn{6}{c}{P-T profile}\\
 $X_\mathrm{H_2O}$ & $X_\mathrm{CH_4}$ & $X_\mathrm{NH_3}$ & $X_\mathrm{CO}$ & $X_\mathrm{HCN}$ & $X_\mathrm{CO_2}$ & $X_\mathrm{C_2H_2}$ & $T_\mathrm{100mb}\mathrm{/K}$    & $\alpha_1\mathrm{/K^{-\frac{1}{2}}}$ & $\alpha_2\mathrm{/K^{-\frac{1}{2}}}$ & $P_1\mathrm{/bar}$ & $P_2\mathrm{/bar}$ & $P_3\mathrm{/bar}$\\
\hline
$5\times10^{-4}$ & $10^{-4}$ & $10^{-4}$ & $10^{-4}$ & $10^{-4}$ & $10^{-4}$ & $10^{-4}$ &$1725$      & $0.42$    & $0.6$ & $0.2$ & $2\times10^{-4}$ & $0.3$      \\
\hline
\end{tabular}
\end{table*}

\subsubsection{P-T Profile Parametrisation}
\label{methods_pt}

Atmospheric temperature profiles can have strong dependence with pressure. Being able to model $P$-$T$ profiles effectively with a minimal number of free parameters is critical, particularly in emission spectroscopy where the spectrum is quite sensitive to the temperature gradient (see fig. \ref{fig:sim_ret_pt}). We adopt the parametric $P$-$T$ profile of \citet{madhu_2009} which is known to be effective in capturing a wide range of $P$-$T$ profiles \citep{madhu_2009,madhu_2011_inv,burningham_2017}. \citet{line_2013} explored alternate parametrisation of the P-T profile, either an analytic profile  \citet{guillot_2010} for grey atmospheres or a level-by-level approach, where the atmosphere is subdivided into several regions and the temperature in each of these layers left as a retrieval parameter. Retrievals with HyDRA were checked with both of these temperature profiles as well to ensure minimal effect of the P-T parametrisation on the derived atmospheric structure, as was seen in \citet{line_2016}.

In the profile we use from \citet{madhu_2009}, the atmosphere is divided into 3 broad regions with the boundaries between them given by $P_0 \leqslant \mathrm{layer~ 1} < P_1$, $P_1 \leqslant \mathrm{layer~ 2} < P_3$ and $P_3 \leqslant \mathrm{layer~3}$. $P_2$ represents the base pressure of the thermal inversion, only present in the atmosphere if $P_1<P_2$. The temperature profile in each layer is given by 
\begin{align}
P &= P_0e^{\alpha_1(T-T_0)^{\beta_1}} \mathrm{~in~layer~1},\\
P &= P_2e^{\alpha_2(T-T_2)^{\beta_2}} \mathrm{~in~layer~2},\\
T &= T_3 \mathrm{~in~layer~3}.
\end{align}
with the free parameters $\alpha_1$, $\alpha_2$, $\beta_1$ and $\beta_2$ determining the gradient of the P-T profile, and $T_i$ representing the temperature at pressure $P_i$. We set $\beta_1=\beta_2=0.5$ as per the reasoning in \citet{madhu_2009}, and fix the top of the atmosphere $P_0 = 10^{-5}$ bar; pressures below this do not significantly affect the observed emission spectrum due to lack of any significant opacity. These conditions, along with the continuity of the temperature between each layer, result in 6 free parameters to fully specify the temperature at any pressure, which for convenience we take to be $T_0$, $\alpha_1$, $\alpha_2$, $P_1$, $P_2$ and $P_3$.

The parameters used in the retrieval are shown in table \ref{tab:params_table}. In the retrieval itself, we choose to parametrise the temperature at 100mb pressure $T_\mathrm{100mb}$ instead of $T_0$. This is convenient as it offers the value of the temperature near the photosphere, and not simply the top of the model atmosphere which would be quite poorly constrained anyway and dependant on the choice of the model (the lowest pressure modelled). It also provides tighter constraints on the other parameters, given that the spectrum is most sensitive to the temperature at this pressure. The measurement at 0.1 bar can also be more easily compared to the planet's equilibrium temperature, and any conclusions we can draw from the observed temperature.

\subsubsection{Chemical Mixing Ratios}
\label{methods_mix_frac}

As well as the P-T structure of the atmosphere, we also wish to determine the abundances of the spectroscopically active species. Molecular species have been discovered in both transmission and emission spectroscopy, most notably H$_2$O, in several transiting exoplanets. As typically considered in retrievals \citep{madhu_2011_inv,line_2016,macdonald_2017} we include molecular species that are expected to be most dominant in H$_2$-rich atmospheres at high temperatures \citep{madhu_2012,moses_2013,heng_lyons_2016}: H$_2$O, CH$_4$, NH$_3$, CO, CO$_2$, HCN and C$_2$H$_2$ (see table \ref{tab:params_table}).

The molecular mixing ratio of species $i$ is given by X$_i = n_i/n_{\mathrm{tot}}$, where $n_i$ is the number density of the species and $n_\mathrm{tot}$ is the total number density of all species. The mixing ratio X$_i$ can be left as a free parameter to be retrieved or fixed to a specific value. The helium fraction $X_{\mathrm{he}}$ is fixed to be 0.15 (the He:H$_2$ ratio is 0.17), i.e solar composition. The hydrogen mixing ratio is calculated from $X_\mathrm{H_2} = 1- \sum_i \mathrm{X}_i$ (only for hydrogen dominated atmospheres), in order to ensure that the sum of the mixing ratios equals unity. Our model takes as input the volume mixing ratios of the molecular species; to convert from the mixing fraction of a species to its ratio relative to H$_2$ it can be multiplied by 1.17.

We consider the atmosphere to be in hydrostatic equilibrium, and the mean molecular weight of the atmosphere $\overline{m}$ is calculated self-consistently using the ideal gas law, taking into account the relative molecular masses of each of the gaseous species $m\mathrm{_i}$ and the corresponding abundance X$_i$. The density and number density $n_\mathrm{tot}$ are given by
\begin{align}
\overline{m} &= \sum_{i} \mathrm{X}_{i}m_{i},\\
P &= \frac{\rho k_b T}{\overline{m}}, \\
n_\mathrm{tot} &= \frac{\rho}{\overline{m}}.
\end{align}

\subsection{Opacities}
\label{methods_opac}

\begin{figure}
  \centering
    \includegraphics[width=\columnwidth,trim=0 0.6cm 0 0,clip]{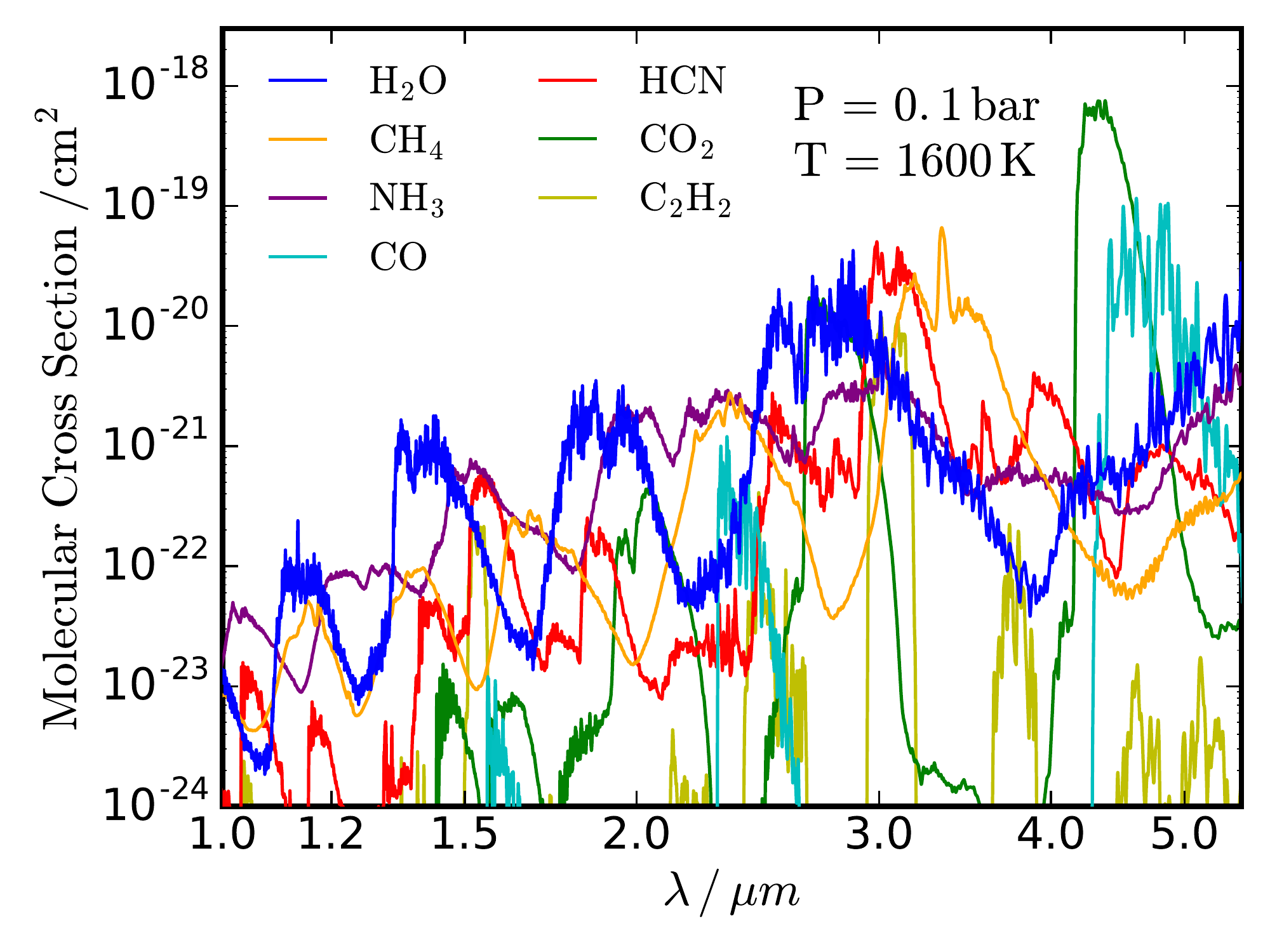}
  \caption{Molecular absorption cross sections for the 7 volatile species considered in our retrievals. These are shown at 1600K, approximately WASP-43b's expected temperature, and at a pressure of 0.1 bar, where most of the emission from the planet originates. The cross sections have been gaussian smoothed in the figure for clarity.}
  \label{fig:cross_sec_compare}
\end{figure}

Spectroscopically active species present can absorb and emit photons in the atmosphere, dependent on the relevant cross-section. These wavelength dependent molecular absorption bands determine the emergent spectrum. Here we describe how the molecular cross-sections are computed for the gaseous species considered, along with the collisionally induced absorption from the dominant H$_2$ and He  gases. The sum of these, the total opacity, is required to calculate the flux ratio from the forward model (section \ref{methods_rad_transfer}). The full numerical details of the computation carried out to compute the cross-sections can be found in \cite{gandhi_2017}. 

In determining the cross-section of each molecule as a function of pressure, temperature and wavelength, we use publicly available molecular ro-vibrational transition line lists for each molecule. In the current version of HyDRA, H$_2$O, CH$_4$, NH$_3$, CO, CO$_2$, HCN and C$_2$H$_2$ molecular absorption is considered. Collisionally induced absorption (CIA) due to H$_2$-H$_2$ and H$_2$-He interactions are also included \citep{richard_2012}. H$_2$O, CO and CO$_2$ are computed from the HITEMP database \citep{rothman_2010}, CH$_4$, NH$_3$ and HCN are calculated from the ExoMol database \citep{tennyson2016} and C$_2$H$_2$ is taken from HITRAN \citep{rothman_2013}. The strength of each transition in the line list is found on a pre-computed grid of temperatures and broadened (with pressure and temperature broadening) to give the cross-section as a function of frequency. These are then summed together for every line and binned down to a 1cm$^{-1}$ spacing grid between 0.4$\micron$ and 50$\micron$. We use a temperature grid with 16 points ranging from 300-3500K and a pressure grid with 7 points evenly spaced in log(P) ranging from $10^{-4}-100$ bar to compute the cross-sections. To obtain the cross-section for a general temperature, pressure and wavelength point, interpolation of the pre-computed grid is carried out. We have tested our models and retrievals for a range of spectral resolutions to ensure minimal differences due to the resolution of our cross section grid, as also demonstrated in \cite{gandhi_2017}.

Once the molecular cross section, $\kappa_{i}(\mathrm{P,T,}\nu)$ is computed for each P-T point of the atmosphere, we simply sum the cross sections of all of the species, weighted by their mixing fractions X$_i$ (see section \ref{methods_mix_frac}). This gives the total extinction coefficient $\chi$ and the optical depth $\tau$ as a function of pressure, temperature and frequency
\begin{align}
\chi(\mathrm{P,T,}\nu) &= \sum_i \mathrm{X}_i n\mathrm{(P,T)} \, \mathrm{\kappa_{i}(\mathrm{P,T,}\nu)} + \sigma_\mathrm{H_2} + \kappa_\mathrm{CIA}, \\
\mathrm{d}\tau(\mathrm{P,T,}\nu) &= \chi(\mathrm{P,T,}\nu) \mathrm{d}z,
\end{align}
where $\sigma_\mathrm{H_2}$ and $\alpha_\mathrm{CIA}$ refer to the loss from the beam due to Rayleigh scattering and collisionally induced absorption of molecular hydrogen respectively. Fig. \ref{fig:cross_sec_compare} shows the molecular cross sections for the 7 molecules considered at 1600K temperature and 0.1 bar pressure, representative of the typical conditions expected in WASP-43b's photosphere.

\subsection{Radiative Transfer}
\label{methods_rad_transfer}
Once the total opacity contribution from each species has been summed, we require calculation of the emergent flux out of the model atmosphere. The radiative transfer method to obtain the emergent planetary spectrum is discussed here. We only consider radiative transfer models in the pure absorption limit, where the scattering into the beam of radiation is assumed to be negligible. However, they begin to deviate from the full solution (i.e ones that account for scattering into the beam) once the wavelength is below $\sim1\micron$ (i.e where scattering can no longer be assumed small). Various methods exist to compute the emergent intensity of radiation, with varying levels of sophistication, but a simple forward model is needed for the retrieval. This is so as to be computationally fast whilst also capturing the relevant physics, as typically retrievals evaluate millions of models to map the parameter space. For this purpose we tested several approaches  in order to determine which would be the most favourable in terms of accuracy and time.

Consider a slab of optical thickness $\tau$ and temperature $T$, with a radiation intensity $I_0$ emergent from underneath at an angle $\theta$ to the normal, with $\mathrm{cos}(\theta) = \mu$. The radiation emergent out of the slab as a function of frequency $\nu$ and angle cosine $\mu$ is given by \citep{seager_2010}
\begin{align}
I_1(\nu,\mu) = I_0(\nu,\mu)e^{-\tau/\mu} + B(T,\nu)(1-e^{-\tau/\mu}),
\end{align}
for a Planck function $B(T,\nu)$ at temperature $T$ and frequency $\nu$. For a model atmosphere with $ND$ layers, we simply compute the contributions of each slab by integrating upwards through the atmosphere along a ray to find the emergent intensity $I_\mathrm{top}$. The flux exiting the top of the atmosphere is given by 
\begin{align}
F_\mathrm{top}(\nu) &= \int_{0}^{2\pi}\int_{0}^{1} \mu I_\mathrm{top}(\nu,\mu) d\mu d\phi
= 2\pi \int_{0}^{1} \mu I_\mathrm{top}(\nu,\mu) d\mu,
\end{align}
where the integral over the azimuth $\phi$ is assumed to be trivial assuming axial symmetry. If the distance to the system is $d$, the observed flux at the observer is then
\begin{align}
F_\mathrm{p}(\nu) = F_\mathrm{top}(\nu)\frac{R_{p,\nu}^2}{d^2}.
\end{align}
Here R$_{p,\nu}$ is the radius of the planetary photosphere (where $\tau_\nu = 1$) at the frequency $\nu$. The mean radius of the planet, required to calculate the temperature profile, is taken to be the observed radius set at a pressure of 0.1 bar which represents the mean pressure of the $\tau_\nu = 1$ surface.

\begin{figure}
	\includegraphics[width=\columnwidth]{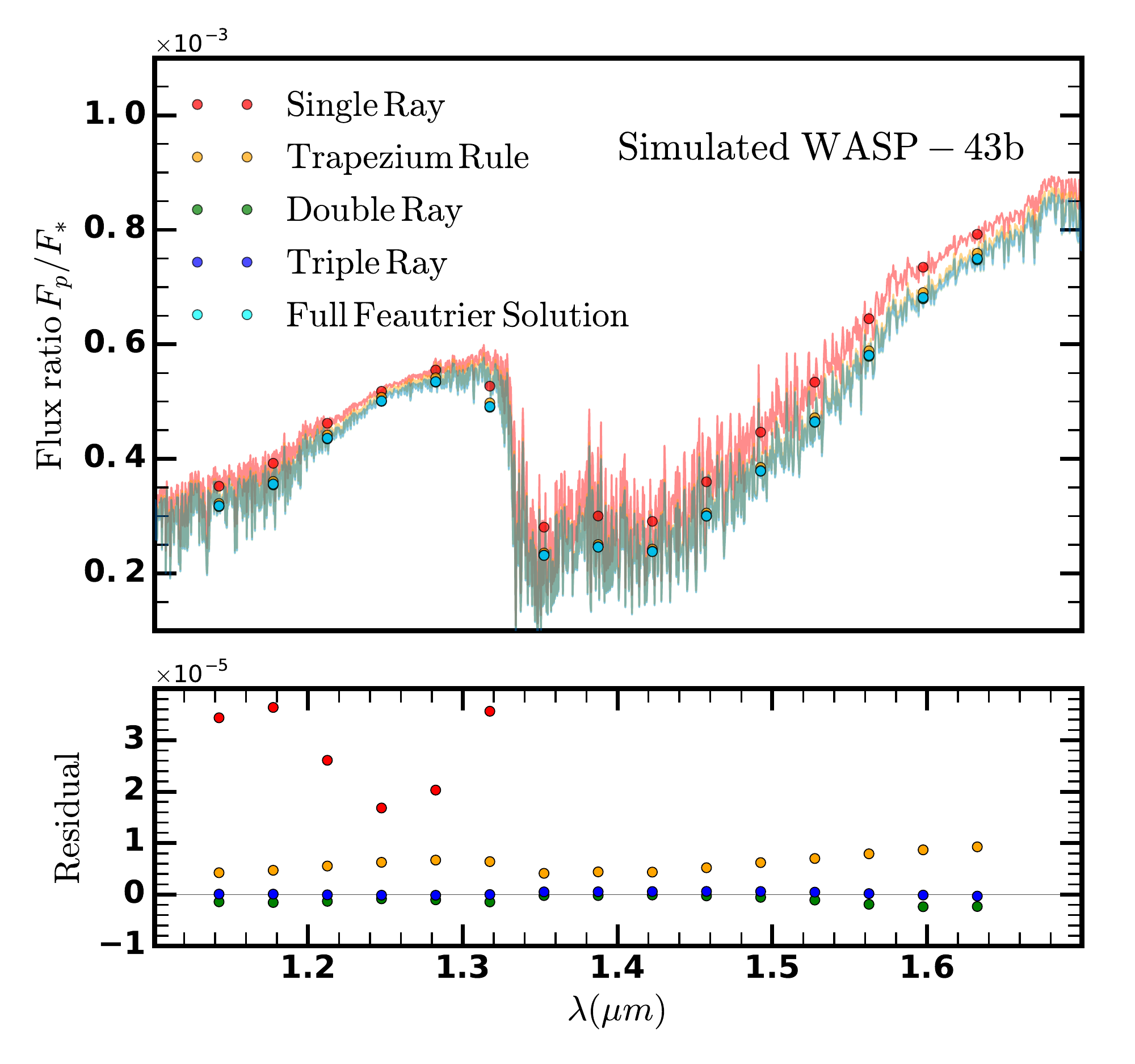}
    \caption{Theoretical emergent flux spectra (top) and residuals (bottom) for various choices of the radiative transfer model shown compared to the full Feautrier radiative transfer solver in \citet{gandhi_2017}. The markers indicate the binned WFC3 spectrum at the resolution of the WASP-43b data. 2000 wavelength points between 1.1 and 1.7$\micron$ were used to generate the spectra with 100 atmospheric layers. The choice of model parameters are given in table \ref{tab:params_table}.}
    \label{fig:fm_compare}
\end{figure}

Our model computes flux from radiation exiting the atmosphere accounting for the loss from the beam due to absorption and scattering. There are multiple approaches to calculate the flux. We experimented with the case where only one ray was considered for the emergent intensity, at $\mu=0$ (vertical), to calculate the flux. We found significant differences ($>30\mathrm{ppm}$) between this and the 6 angle dependent radiative transfer using the Feautrier method as described in \cite{gandhi_2017} (see fig. \ref{fig:fm_compare}). This single ray approach is the method used in \cite{line_2013} and \cite{madhu_2009}, and has been reasonable given previous observational data. It is computationally fast, requiring only one ray's exiting radiation field to be calculated. Perhaps a more representative angle could have been used for single ray calculations, e.g $\mu=1/\sqrt[]{3}$. However, given the high accuracy on the WASP-43b dataset, we chose to explore other methods in order to calculate the flux to a greater accuracy. This is achieved by integrating over multiple angles. Trapezium rule integration over $\mu$ with 6 evenly spaced values of $\mu$ ranging between 0 and 1 significantly improves the result, as the variation of the intensity with angle is considered. In this latter case, the difference from the full Feautrier calculation is below 10 ppm as shown in fig. \ref{fig:fm_compare}. However, just two or three angles with Gaussian quadrature provided excellent match to the full solution (see table. \ref{tab:quadrature_weights}). Increasing the number of angles further did not result in major differences in the spectra, but did increase the computation time significantly (time per model scales with the number of $\mu$ values). It is for this reason that we adopt two angles with Gaussian quadrature in our model. For comparison, the precision of the WASP-43b data in secondary eclipse considered here is $\sim35$ppm, and so the present accuracy in our radiative transfer more than suffices for our purpose. In future, however, with higher quality data sets it would be imperative to use as accurate a solution as possible to ensure accurate constraints.

\begin{table*}
	\centering
	\caption{Weights and angles ($\mu$) used for each integration scheme. The angles for the integration with the Feautrier method can be found in \citet{gandhi_2017}, but are the same as triple ray quadrature.}
	\label{tab:quadrature_weights}
\begin{tabular}{llc}
\hline
Method & Weights & $\mu$ \\
\hline
Single Ray & 1 & 1 \\
Trapezium Rule & 0.1, 0.2, 0.2, 0.2, 0.2, 0.1 & 0, 0.2, 0.4, 0.6, 0.8, 1 \\
Double Ray &  0.5, 0.5 & $\frac{1}{2}-\frac{1}{2}\sqrt[]{\frac{1}{3}}$,   $\frac{1}{2}+\frac{1}{2}\sqrt[]{\frac{1}{3}}$ \\
Triple Ray & 5/18, 4/9, 5/18 & $\frac{1}{2}-\frac{1}{2}\sqrt[]{\frac{3}{5}}$,   $\frac{1}{2}$,   $\frac{1}{2}+\frac{1}{2}\sqrt[]{\frac{3}{5}}$ \\

\end{tabular}
\end{table*}

\subsection{Stellar Spectrum}
\label{stellar_spectrum}
The data sets presented in emission spectroscopy of transiting hot Jupiters are given in terms of the ratio of the planetary to stellar flux (e.g see fig. \ref{fig:fm_compare}). It is therefore vital that in order to obtain an accurate spectrum that we compute the flux of the host star. We calculate this using the available properties of the star and the Kurucz model grid of spectra \citep{Kurucz_1979_paper,kurucz_model}. The effective temperature of the star, the surface gravity, log(g), and metallicity are interpolated on the model grid to obtain the flux $F_\mathrm{top,star}$ at the top of the stellar atmosphere. The observed flux at earth is then given by
\begin{align}
F_\mathrm{star}(\nu) = F_\mathrm{top,star}(\nu)\frac{R_\mathrm{star}^2}{d^2}.
\end{align}
We then use this flux along with $F_\mathrm{p}$ from section \ref{methods_rad_transfer} to calculate the theoretical data points for such a model below. We also checked against a simple Planck function for the star's flux, and did see some small differences where the stellar spectrum had absorption features or varied significantly from the Planckian. This was as expected and when the noise constraints on the data set was greater than 20 ppm in the WFC3 bandpass as in our case the differences were negligible. It is however noteworthy that the quality of the stellar spectrum will be important to interpret exoplanetary  spectra with very high precision. 

\subsection{Generating Model Data}
\label{methods_bin_data}
The spectroscopic instruments used to obtain the model data points for a given spectrum (e.g Hubble's WFC3 and Spitzer's IRAC) have their own transmission properties which must be taken into account when generating the model data points. All of the instruments have their own sensitivity as a function of wavelength; the Hubble WFC3 bandpass also has multiple grisms that results in the received flux being convolved with an instrument point spread function ($PSF$) before it is binned (in our case only the G141 grism requires this). Hence, to accurately determine where the spectral points for our model spectrum lie, a given theoretical flux $F$ is firstly convolved where necessary to give
\begin{align}
F_\mathrm{conv}(\lambda) = \int_{-\infty}^{\infty} F(\lambda') PSF(\lambda-\lambda') d\lambda'.
\end{align}
Conversion from frequency space, $F(\nu)$, to wavelength space, $F(\lambda)$, is easily given by
\begin{align}
F(\lambda) = F(\nu) \mid \frac{d\nu}{d\lambda}\mid = F(\nu) \frac{c}{\lambda^2}.
\end{align}
The convolved spectrum $F_\mathrm{conv}$ is then multiplied by the instrument sensitivity function $S(\lambda)$ and normalised to obtain the binned flux,
\begin{align}
F_\mathrm{binned} = \frac{\int_{\lambda_\mathrm{min}}^{\lambda_\mathrm{max}} F_\mathrm{conv}(\lambda)S(\lambda) d\lambda}{\int_{\lambda_\mathrm{min}}^{\lambda_\mathrm{max}} S(\lambda) d\lambda},
\end{align}
with the bin edges having a minimum and maximum wavelength $\lambda_\mathrm{min}$ and $\lambda_\mathrm{max}$ respectively. The model spectrum needs to be convolved with the $PSF$ only for the HST WFC3 spectrograph and not for the Spitzer photometric bands. These steps are carried out for the stellar and the planetary fluxes $F_\mathrm{star}$ and $F_\mathrm{p}$ and the final model points (given as a flux ratio) are
\begin{align}
y_\mathrm{model,n} = \frac{F_\mathrm{p,binned,n}}{F_\mathrm{star,binned,n}},
\end{align}
for every bin n that is considered.

\subsection{Parameter Estimation and Statistical Inference}
\label{multinest}

In atmospheric retrieval, a parametric forward model is coupled to a statistical inference algorithm to estimate the model parameters given the data and to perform model comparisons. Contemporary retrieval codes routinely use rigorous Bayesian statistical inference methods such as MCMC \citep{madhu_2011_inv,line_2013} and nested sampling \citep{macdonald_2017,benneke_2013,line2_2016,lavie_2017}. We utilise nested sampling \citep{skilling_2004}, which is advantageous in that it allows for calculation of the Bayesian evidence and hence model comparisons.

In our work, we employ the multimodal nested sampling algorithm MultiNest \citep{feroz_2008,feroz_2009,feroz_2013}, using the python package PyMultiNest developed by \cite{buchner_2014}. The full details of the statistical methods and the priors used for the analysis can be found in \cite{macdonald_2017}. We briefly summarise the approach below for convenience. We begin with the statistical techniques used to ascertain how well the model describes observations, and proceed afterwards onto the Bayes factor calculations to compare different models.

\subsubsection{Bayesian Evidence}
\label{bayes_evidence}
Given a set of parameters $\theta$ that describe some forward models $M_k$, we have a set of a priori expectations on the values through a prior probability density function $\pi(\theta,M_k)$. $p(\mathbf{\theta}|{y_\mathrm{obs}},{M_k})$ is the prior posterior probability distribution. With the observations and spectral data points $\mathbf{y_\mathrm{obs}}$ and $\mathbf{y_\mathrm{k}}$ respectively, this can be rewritten \citep{trotta_2008} utilising Bayes theorem to
\begin{align}
p(\mathbf{\theta}|{y_\mathrm{obs}},{M_k}) &= \frac{\mathcal{L}({y_\mathrm{obs}}|\theta , M_k) \pi (\theta, M_k)}{\mathcal{Z}({y_\mathrm{obs}}|{M_k})},
\end{align}
where we have defined the likelihood function, prior and the Bayesian evidence as $\mathcal{L},~\pi$ and $\mathcal{Z}$ respectively. We will take the likelihood function to be
\begin{align}
\mathcal{L}({y_\mathbf{obs}}|\theta, M_k) = \prod_{i}^{N_\mathrm{obs}} \frac{1}{\sqrt[]{2\pi}\sigma_i} \mathrm{exp}\bigg(\frac{-(y_\mathrm{obs,i} - y_\mathrm{k,i})^2}{2\sigma_i^2}\bigg).
\label{eqn:likelihood}
\end{align}
This is a measure of how likely the choice of spectrum produces the observed data points, a higher likelihood indicates that the set of parameters is favoured. Eqn. \ref{eqn:likelihood} assumes that the error is independently gaussian distributed for each data point. To avoid bias, uniform priors are often used. The Bayesian evidence $\mathcal{Z_\mathrm{k}}$ for a spectrum $\mathrm{k}$ is simply a normalisation factor and is given by an integral over all parameter space
\begin{align}
\mathcal{Z_\mathrm{k}} = \int_{\theta} \mathcal{L}({y_\mathrm{obs}}|\theta , M_k) \pi (\theta, M_k) d\theta
\end{align}
The Bayesian evidence is simply a ``figure of merit" that assesses the ability of a given model to descrive the data and can hence be used for model comparisons (see below).

\subsubsection{Bayes Factor}
\label{bayes_factor}
Whilst comparing two models M$_0$ and M$_1$, we require the Bayes factor
\begin{align}
B_{01} \equiv \frac{\mathcal{Z}({y_\mathrm{obs}}|{M_0})}{\mathcal{Z}({y_\mathrm{obs}}|{M_1})}.
\end{align}
This describes how one model performs relative to another in explaining the observations. If $B_{01}>1$, model M$_0$ is favoured over M$_1$ given the observed data (its Bayesian evidence is greater). This allows us to demonstrate quantitatively whether a more complex model or a simple one is required. For instance, if we wish to calculate the significance of a detection of a molecule, we can find the Bayes factor $B_{01}$ for a model with the molecule (M$_0$) and one with the same parameters but with this molecule removed (M$_1$). This then establishes the evidence for such a molecule. Bayes factors greater than 3, 12 and 150 are often quoted as weak, moderate and strong detections. The larger the Bayes factor, the more evidence there is to support model M$_0$, i.e the better the explanation of the data with the relevant parameter. This can then also be converted into a detection significance. For further details on Bayesian analysis, we refer the reader to \cite{trotta_2008}. The detection significances can tell us not only about the species that is present but given equilibrium models also provide constraints on disequilibrium phenomena.

\subsection{Constraints on Disequilibrium}
\label{methods_diseqm}

One of the key functionalities of HyDRA is to constrain the deviations of the retrieved compositions and $P$-$T$ profiles from chemical and radiative-convective equilibrium. This is achieved by operating the retrieval code in tandem with our self-consistent equilibrium model for exoplanetary atmospheres GENESIS \citep{gandhi_2017}. As modules from both the retrieval and equilibrium models are shared, any deviations between retrieved and equilibrium properties are unlikely to be due to the intrinsic model differences but rather to non-equilibrium atmospheric processes at play.

\subsubsection{Radiative-Convective Disequilibrium}
\label{methods_rad_diseqm}
We constrain radiative-convective disequilibrium by investigating deviations of retrieved $P$-$T$ profiles from equilibrium $P$-$T$ profiles with the same retrieved compositions. This is done by considering the posterior distributions of the retrieved $P$-$T$ profiles, and their corresponding compositions, and computing the equilibrium $P$-$T$ profiles using GENESIS \citep{gandhi_2017} by keeping the compositions fixed to the retrieved values. The resultant equilibrium $P$-$T$ profiles are used to compute the temperature differential ($\Delta$T) between the equilibrium and retrieved profiles as a function of altitude (or pressure). The profile of $\Delta$T and its statistical uncertainties provides a measure of the deviation from radiative-convective equilibrium. 

The GENESIS model determines the atmospheric P-T profile in equilibrium, such that the incoming and outgoing radiation from each layer in the atmosphere is equal. The condition of radiative-convective equilibrium in each layer is given by (see \citealt{gandhi_2017} for full details)
\begin{align}
\int_0^\infty \kappa_\nu (J_\nu-B_\nu) d\nu + \frac{\rho g}{4\pi}\frac{dF_\mathrm{conv}}{dP} & = 0.
\end{align}
Here, $\kappa_\nu$ refers to the absorption coefficient, $J_\nu$ is the mean radiation intensity and $B_\nu$ is the Planck function at a frequency $\nu$, and the convective flux $F_\mathrm{conv}$ is applied where convective regions occur in the atmosphere. Whereas the retrieval's P-T profile is parametrised, the GENESIS model calculates the temperature for every pressure layer in the model atmosphere. The opacity and stellar flux calculations are identical to the retrieval algorithm, and the full radiative transfer scheme shows negligible differences as explored in section \ref{methods_rad_transfer}. 

As we hold the chemistry fixed to a range of the retrieved values, we know that any deviation that arises can be attributed to effects not accounted for. These differences can provide key insights into atmospheric energy transport, and about the validity of 1-D models. It also provides clues to the next steps in our modelling approaches in order to match observations and incorporate new physics, e.g atmospheric dynamics, into 1-D models used in retrievals.

\subsubsection{Chemical Disequilibrium}
\label{methods_chem_diseqm}
We constrain chemical disequilibrium in the atmosphere by considering deviations of the retrieved compositions from those computed assuming chemical equilibrium for the same retrieved $P$-$T$ profiles. Given the posterior distributions of the retrieved parameters, we choose a representative statistical sample of compositions and their corresponding $P$-$T$ profiles. These $P$-$T$ profiles are then used to compute equilibrium abundances of all the chemical species and the corresponding uncertainties. The differences between the retrieved and equilibrium abundance provide constraints on the deviations from chemical equilibrium. The species include H$_2$O, CH$_4$, NH$_3$, CO, HCN, CO$_2$, C$_2$H$_2$, C$_2$H$_4$ and N$_2$. To determine the equilibrium mixing fractions for a species at a given pressure and temperature, we utilise the semi-analytic calculations developed by \cite{heng_2016}. The main chemical equations that govern the mixing ratios in a hydrogen dominated atmosphere are
\begin{equation}
\begin{split}
\mbox{CH}_4 + \mbox{H}_2\mbox{O} &\leftrightarrows \mbox{CO} + 3\mbox{H}_2 , \\
\mbox{CO}_2 + \mbox{H}_2 &\leftrightarrows \mbox{CO} + \mbox{H}_2\mbox{O},\\
2\mbox{CH}_4 &\leftrightarrows \mbox{C}_2\mbox{H}_2 + 3\mbox{H}_2, \\
\mbox{C}_2\mbox{H}_4 &\leftrightarrows \mbox{C}_2\mbox{H}_2 + \mbox{H}_2, \\
2\mbox{NH}_3 &\leftrightarrows \mbox{N}_2 + 3\mbox{H}_2,\\
\mbox{NH}_3 + \mbox{CH}_4 &\leftrightarrows \mbox{HCN} + 3\mbox{H}_2.\\
\end{split}\label{eqn:chem_eqns}
\end{equation}
Given the elemental abundances relative to atomic hydrogen for carbon, oxygen and nitrogen (which we take as solar, a reasonable assumption given that the retrieved water abundance is consistent with solar, see section \ref{results}), the abundances of each of the species in the set of chemical equations above can be determined. This is achieved by solving a decic equation (see \citealt{heng_2016}) knowing each of the equilibrium constants, which are functions of pressure of temperature. Therefore, for a fixed P-T profile we can determine the mixing ratios for our hydrogen dominated atmosphere. For the purpose of this work, we ignore N$_2$ as it offers no spectral signature, as well as C$_2$H$_4$ as it is present only in very small quantities and thus does not significantly affect the spectra and cannot be retrieved to any certainty for our spectral range.  

\section{Retrievals with Simulated Data}
\label{validation}

\begin{figure*}
\begin{overpic}[width=\textwidth]{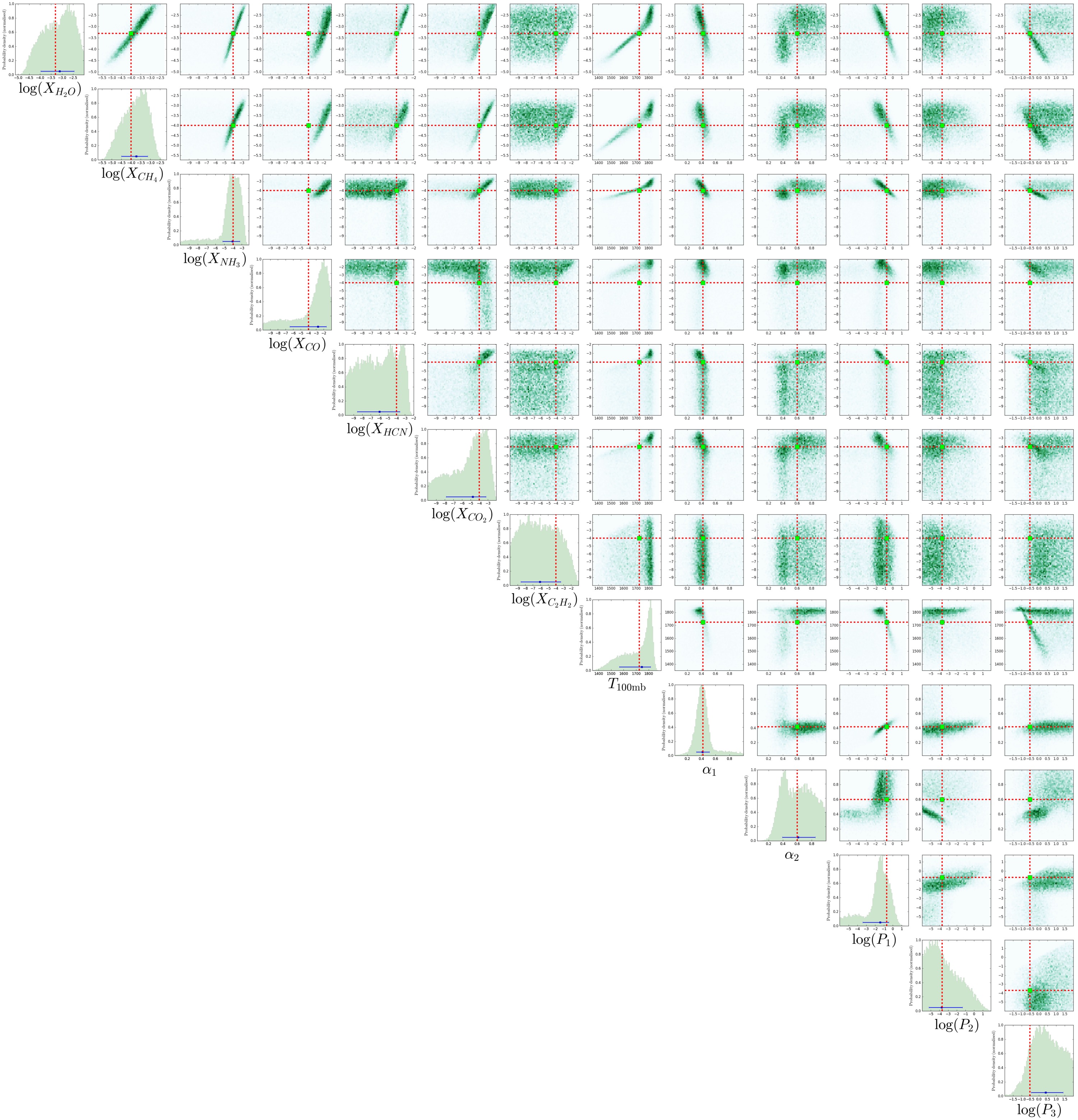}
\put (0,37) {\begin{tabular}{l*2{c}r}
Parameter              & Actual & Retrieved & Error \\
\hline
log(X$_\mathrm{H_2O}$) & -3.3 &-3.14 & $\substack{+0.59 \\ -0.71}$ \\
log(X$_\mathrm{CH_4}$) & -4.0 &-3.74 & $\substack{+0.50 \\ -0.62}$ \\
log(X$_\mathrm{NH_3}$) & -4.0 &-4.08 & $\substack{+0.70 \\ -1.29}$ \\
log(X$_\mathrm{CO  }$) & -4.0 &-2.8 & $\substack{+1.1 \\ -3.7}$ \\
log(X$_\mathrm{HCN }$) & -4.0 & < -2.8 & -\\
log(X$_\mathrm{CO_2}$) & -4.0 &< -2.5& -\\
log(X$_\mathrm{C_2H_2}$)& -4.0 & < -1.9 & - \\
T$_\mathrm{100mb}$ /K   & 1725 &1730 & $\substack{+77 \\ -142}$\\
$\alpha_1 \, /\mathrm{K}^{-\frac{1}{2}}$     & 0.42 & 0.43 & $\substack{+0.11 \\ -0.09}$\\
$\alpha_2 \, /\mathrm{K}^{-\frac{1}{2}}$     & 0.6 &0.59 &$\substack{+0.25 \\ -0.20}$\\
log(P$_1$/bar)     & -0.70 &-1.5 & $\substack{+1.0 \\ -2.3}$\\
log(P$_2$/bar)     & -3.70 &-3.9 & $\substack{+2.3 \\ -1.4}$\\
log(P$_3$/bar)     & -0.52 &0.37 & $\substack{+1.00 \\ -0.82}$\\
\end{tabular}
}
\end{overpic}
    \caption{Marginalised posterior distribution for the synthetic retrieval on the emergent dayside spectrum of WASP-43b, with the simulated spectrum taken from our self-consistent model GENESIS \citep{gandhi_2017} so as to be in radiative equilibrium. The red lines indicate the actual value of each parameter in the posterior corner plot and the histograms show the retrieved values and their error. The blue error bars indicate the median and 1$\sigma$ error bars. Over $10^6$ models were run with 4000 live samples and 4000 wavelength points between 1 and 5.5$\micron$, with the model atmosphere consisting of 100 atmospheric layers. The table on the left shows the actual and retrieved parameters along with their associated uncertainty. The P-T profile parameters are described in section \ref{methods_pt} and the abundances and opacity calculations are described in section \ref{methods_opac}.}
    \label{fig:sim_ret_corner}
\end{figure*}

\begin{figure}
	\includegraphics[width=\columnwidth]{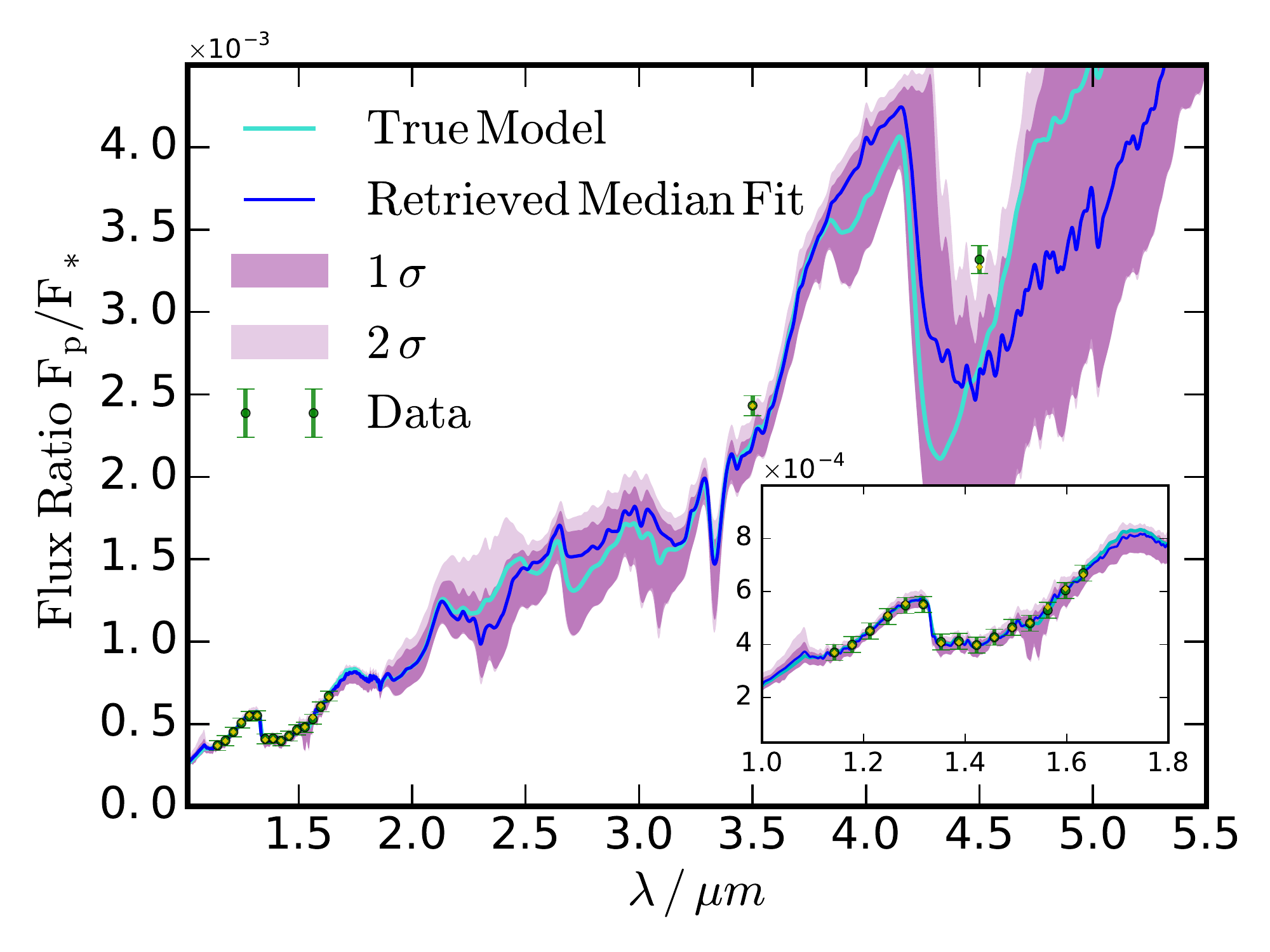}
    \caption{Retrieved emergent spectrum from the simulated data of WASP-43b. The spectrum used for our retrieval was taken from the self-consistent model shown in cyan. The best-fitting model is shown in blue along with its binned data points as yellow diamonds. The dark and light purple contours show the 1$\sigma$ and 2$\sigma$ spread of 4000 parameter combinations from the posterior. The green markers indicate the spectral data points for the WFC3 (see inset) and Spitzer 3.6$\micron$ and 4.5$\micron$ channels and their associated error. The noise on the data is identical to the actual spectrum for the planet.}
    \label{fig:sim_ret_spectrum}
\end{figure}

\begin{figure}
	\includegraphics[width=\columnwidth]{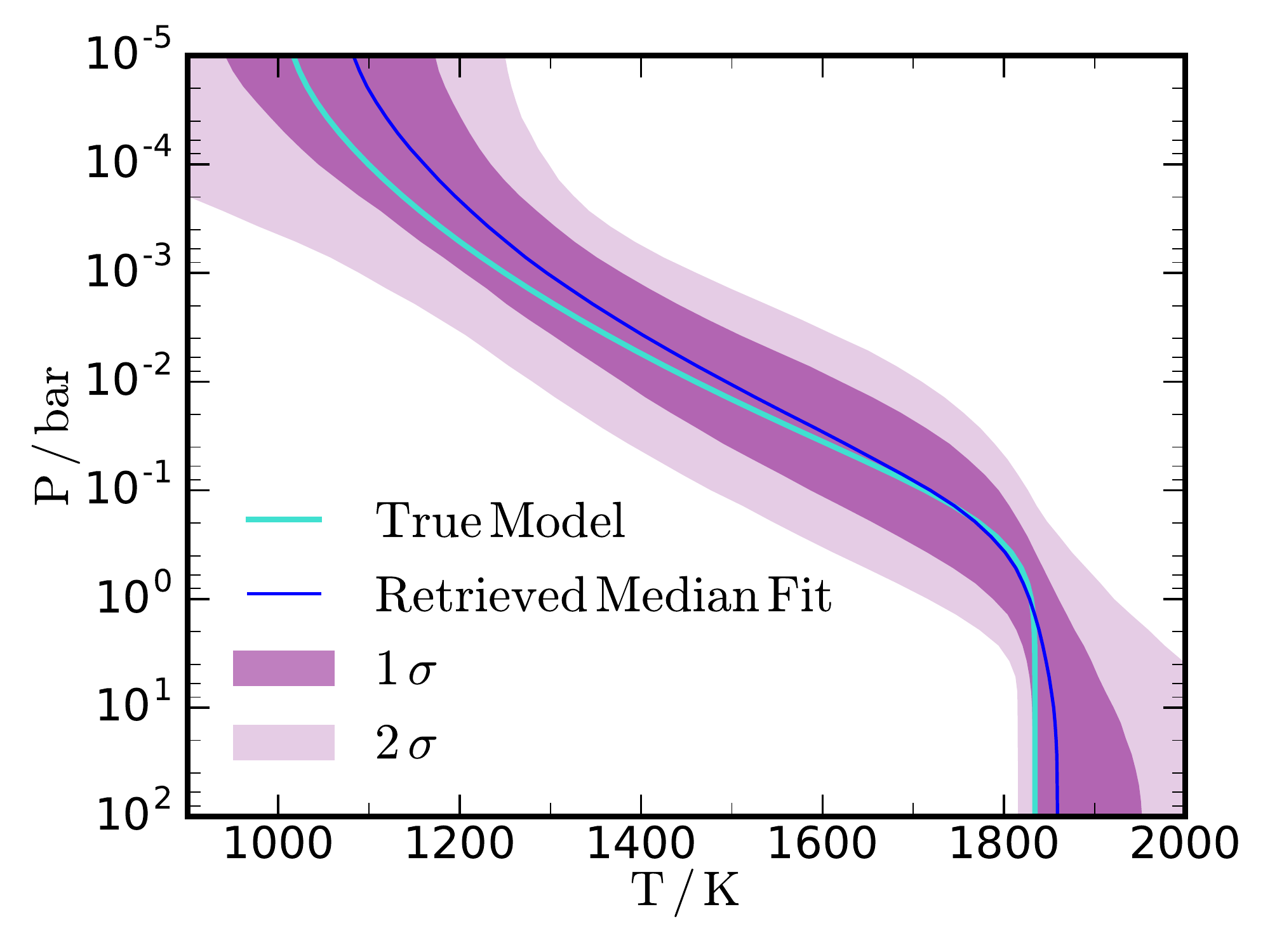}
    \caption{Retrieved dayside P-T profile from the simulated spectrum of WASP-43b, with the actual P-T profile used to generate the simulated data shown in cyan. The blue line indicates the median fit and the dark and light purple contours show the 1$\sigma$ and 2$\sigma$ spread in the results drawn from the posterior.}
    \label{fig:sim_ret_pt}
\end{figure}

We now proceed to validate the HyDRA retrieval framework. The aim of this exercise is to demonstrate the effectiveness of HyDRA for atmospheric retrieval with emission spectra and to explore any degeneracies that may be present in the model parameter space given a dataset. We apply the code to synthetic data to demonstrate its effectiveness in extracting parameter values that are known a priori. Each component of HyDRA, i.e the P-T parametrisation, the opacity calculations and the simplified radiative transfer module were also checked against a fully self-consistent radiative-convective equilibrium model (see fig. \ref{fig:fm_compare}) to ensure a common modelling framework.

We apply HyDRA to a simulated thermal emission spectrum of the hot Jupiter WASP-43b. The choice of this system was driven by the quality of data available and its conduciveness for thermal emission observations. WASP-43b is a hot Jupiter with a mass of $\sim$2 M$_J$ and radius of 0.93 $R_J$ and orbits a K7 dwarf \citep{hellier_2011}. The relatively small and cool star leads to a high planet-star flux ratio making the planet a prime target for thermal emission observations. Consequently, WASP-43b is the most observed transiting planet in thermal emission with high S/N observations using HST WFC3 \citep{stevenson_2014} and Spitzer \citep{blecic_2014}. The synthetic emission spectrum was generated using a theoretical spectrum computed using our GENESIS self-consistent radiative-convective equilibrium model \citep{gandhi_2017}. The compositions for the volatile atmospheric molecules shown in table \ref{tab:params_table}. The model atmosphere is in radiative-convective but not thermochemical equilibrium, to demonstrate the retrieval's ability to constrain the chemical species when they are out of chemical equilibrium. Many of the modules between the models are shared (see section \ref{methods_rad_diseqm}). 

We have performed consistency and sensitivity analyses of our modelling framework over a range of fixed parameters and adopt the most optimal values. The model spectrum is binned to obtain simulated data points with the same precision and resolution as those obtained from real observations in the WFC3 and  Spitzer IRAC bandpasses \citep{kreidberg_2014,blecic_2014}. The parameters of the synthetic model are shown in table \ref{tab:params_table}, with the chemical abundance fixed for all pressures for consistency with the retrieval framework. The uncertainties in the simulated spectrum are taken to be exactly the same as those of the observed spectrum discussed above and added as random gaussian noise on the synthetic model. The parameters in the retrieval include abundances of the seven prominent volatile species expected in hot Jupiters (H$_2$O, CH$_4$, NH$_3$, CO, HCN, CO$_2$ and C$_2$H$_2$) and six parameters that describe the P-T profile. The spectral resolution of the model was set to contain 4000 points uniformly spaced between 1-5.5 $\micron$, for both the self-consistent model as well as retrievals. The retrievals were carried out using the nested sampling algorithm with 4000 live sample points. We have considered a range of spectral resolutions, number of live sample points in nested sampling, variation in pressure grid in the $P$-$T$ profiles, etc, and have adopted the most conservative values for each. We also tested the radiative transfer with a simple single ray $\mu=0$ case against integration over multiple angles (see section \ref{methods_rad_transfer}) and have adopted the optimal two-angle quadrature. In what follows, we discuss the results of our retrievals with the simulated data, given in table \ref{tab:params_table}. 

\subsection{Retrieved Abundances}

The retrieved abundances show very good agreement with the true abundances of the synthetic model. Figure~\ref{fig:sim_ret_corner} shows the posterior distributions for the retrieved chemical species along with their estimated and true values. We find that the true values lie within the 1-$\sigma$ uncertainties for all the species. The best retrieved molecule is H$_2$O with the median value only 0.16 dex away from the true value and an average uncertainty of 0.65 dex. Other molecules retrieved at similar precision include CH$_4$ and NH$_3$. The detection significances of H$_2$O, CH$_4$, and NH$_3$ obtained from nested model comparisons are 6.7$\sigma$, 5.6$\sigma$, and 2.1$\sigma$, respectively. The reason behind the well constrained abundances of these three molecules is a combination of their high abundances used in the synthetic model as well as their relatively strong spectral features mainly in the WFC3 bandpass; H$_2$O being the strongest and NH$_3$ the weakest of the three. The lower significance of NH$_3$ is also due to degeneracies with HCN in the WFC3 bandpass as discussed below \citep[also see][]{macdonald_2017}.

The other molecules only have weak constraints, largely owing to the lack of significant unique features in the observed spectral range. The spectral features of the molecules in the observed wavelength range are shown in fig.~\ref{fig:cross_sec_compare}. HCN has a minor feature at $\sim1.55\micron$, mostly overwhelmed by the NH$_3$ and H$_2$O \citep{macdonald_2017} and as such allows for only an upper-limit of $\mathrm{X_{HCN}}<10^{-2.8}$ that is still consistent with the true value. The C$_2$H$_2$ has very weak features so there is almost no constraint on its abundance $\mathrm{X_{C_2H_2}}<10^{-2}$. In principle, CO and CO$_2$ have strong features in the Spitzer 4.5 $\micron$ IRAC bandpass but are degenerate with each other for the same reason, as discussed below. As such their abundances are also relatively unconstrained. Finally, the spectral resolution and precision of the data also limit the capability of the retrieval in constraining the abundances. 

\subsubsection{CO$_2$ Degeneracy}
\label{co2_degeneracy}

As alluded to above we find CO and CO$_2$ to be degenerate given the single IRAC 4.5 $\micron$ band with significant features. The CO and CO$_2$ abundances show an ``L"-shaped correlation, as also seen in previous studies \cite{madhu_2011_inv, line_2016}. The absorption in the 4.5 $\micron$ band can be explained by either a high CO and low CO$_2$ or vice versa or a combination of both, leading to the L-shaped behaviour. The weak constraints on the species are  also contributed by the large uncertainty in the 4.5 $\micron$ Spitzer IRAC data point. The detection significance of CO or CO$_2$ individually is weak, but the joint detection significance of having either one of CO or CO$_2$ is over 10$\sigma$. With our chemical equilibrium model \citep{gandhi_2017} and full equilibrium calculations for similar planets \citep{moses_2011,moses_2013}, we find that the H$_2$O abundance always exceeds the CO$_2$ abundance, regardless of the C/O ratio and metallicity. Hence, when considering hot Jupiter retrievals, we impose the constraint that the H$_2$O abundance must exceed the CO$_2$ abundance, as suggested in previous studies\citep{madhu_2012,heng_lyons_2016}. This partially breaks the degeneracy and allows for more chemically realistic mixing fractions. It should be noted that this assumption can only be made when considering H$_2$ rich atmospheres. This allows for tighter constraints on the other species, and the temperature can be retrieved more accurately.

\subsubsection{Abundance-T$_\mathrm{100mb}$ Degeneracy}
\label{h2o_degeneracy}

The molecular abundances are generally correlated with the retrieved temperature (see fig. \ref{fig:sim_ret_corner}). A higher abundance raises the opacity, and hence emission is seen to occur from higher in the atmosphere. This increase in the chemical abundance can be compensated by a subsequent alteration in the temperature profile, as long as the temperature at which the emission occurs in the atmosphere remains the same. Given the observational uncertainties in our dataset, we find the abundances and T$_\mathrm{100mb}$ to lie on a line of degeneracy (see fig. \ref{fig:sim_ret_corner}). We see from fig. \ref{fig:sim_ret_spectrum} that in the WFC3 bandpass there is no significant deviation between the retrieved and true models. However, as the spectrum proceeds away from the WFC3 spectral range the uncertainty in the retrieved flux ratio increases due to the lack of high precision data. Observations at other wavelengths in future with instruments such as JWST may be able to better constrain the flux ratio and hence the atmospheric properties more effectively. The Spitzer IRAC photometric points at 4.5, 5.8 and 8 $\micron$ channels may also be used to address the problem, however, these points often come with larger associated errors and difficult systematics \citep{diamond-lowe_2014} making it difficult to do so. 

\subsection{Retrieved P-T Profile}

Similar to the chemical abundances the parameters of the $P$-$T$ profile are also retrieved at high accuracy. The retrieved posterior distributions of the parameters and their estimated values are shown in Figure~\ref{fig:sim_ret_corner}. The true values of all the parameters are recovered to within the 1-$\sigma$ uncertainties. The retrieved temperature profile along with the confidence limits is shown in fig. \ref{fig:sim_ret_pt}, and is in agreement with the true profile within 1$\sigma$ for almost all pressures. The fit is particularly good at $\simeq$1mbar-1 bar, with an error of only $\lesssim$50 K on average between the median profile and the true value. The photospheric temperature is also well constrained with an average uncertainty on T$_{\rm 100 mb}$ to be $\sim$110 K. The retrieval captures the temperature gradient near the photosphere accurately, unsurprising as this is where the bulk of the spectrum is generated. Naturally, the constraints on the temperature are weaker in the lowest and highest regions of the atmosphere that are inaccessible to the observations. 

\subsection{Radiative-Convective Disequilibrium}

We now use the retrieved posterior distributions of the $P$-$T$ profiles to constrain deviations from radiative-convective equilibrium. This is done by considering the chemical compositions corresponding to the retrieved $P$-$T$ profiles and deriving the corresponding $P$-$T$ profiles in radiative-convective equilibrium using the GENESIS self-consistent model (see section \ref{methods_rad_diseqm}). This provides a test of our model's ability to infer disequilibrium in the temperature, as the simulated data set used for the retrieval was in thermal equilibrium. The GENESIS model was run for 1000 randomly sampled retrieved chemical abundances from the posterior, and calculating the equilibrium P-T profile this chemistry. 

\begin{figure*}
	\includegraphics[width=\textwidth]{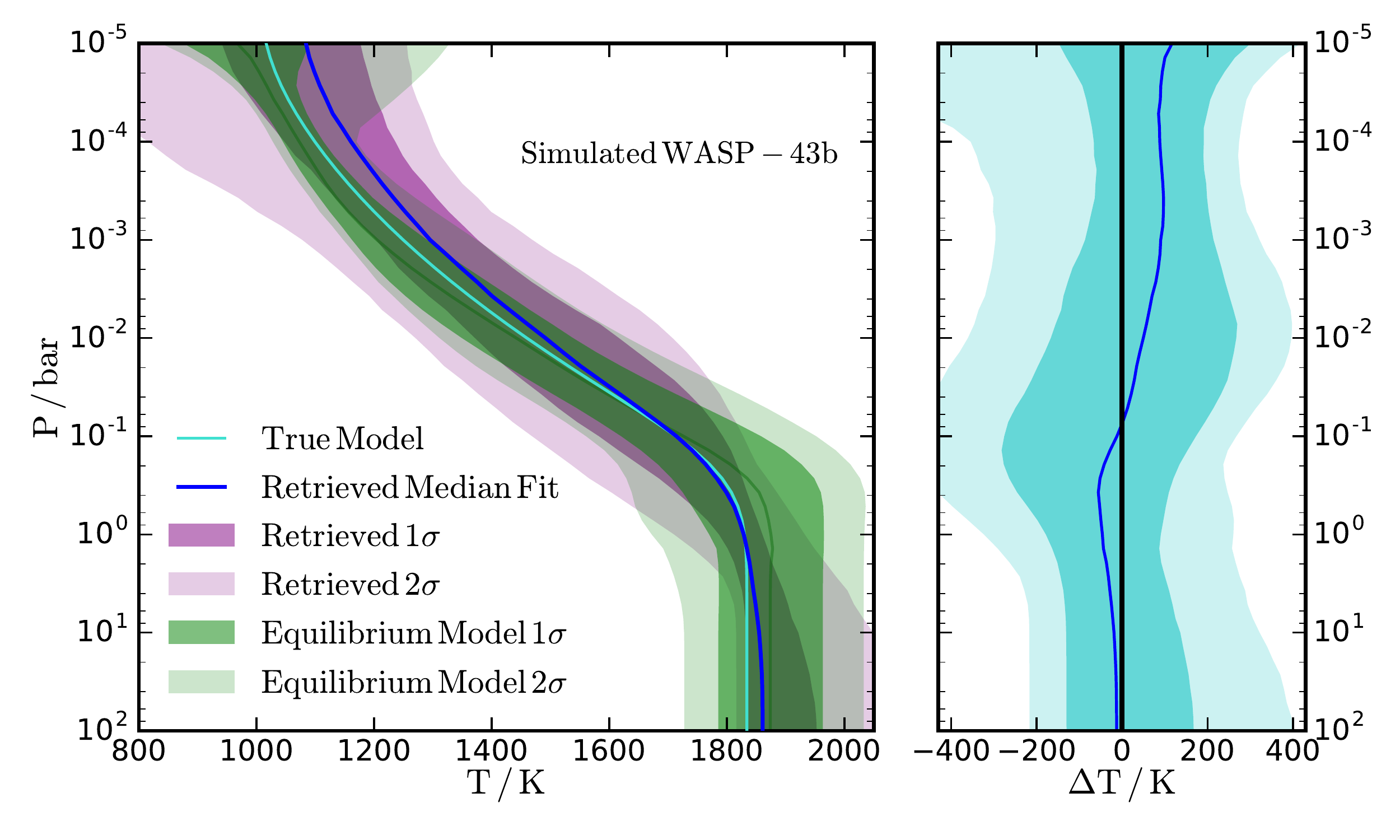}
    \caption{Retrieved P-T profiles (purple) and theoretical radiative equilibrium calculations performed with GENESIS \citep{gandhi_2017} (in green) for the dayside of WASP-43b shown on the left. The cyan line indicates the P-T profile of the simulated data. The right hand side shows the temperature difference between the median retrieved and equilibrium models, and the associated 1 and 2 $\sigma$ confidence contours. The retrieval's solution was used to run the equilibrium model with retrieved chemistry, to determine the radiative-convective profile and hence the equilibrium profile. The spread in the equilibrium model's P-T profile is due to the chemical variations in the retrieval's solution. The sodium and potassium abundances were set to solar compositions for the equilibrium calculations.}
    \label{fig:sim_diseqm_rc}
\end{figure*}

Fig. \ref{fig:sim_diseqm_rc} shows both the retrieved and the radiative-equilibrium P-T profiles, along with the actual P-T profile of the simulated data. All lie within 1$\sigma$ of each other, particularly in the vicinity of $\tau_\nu=1$. Hence with the combination of our self-consistent and our retrieval algorithm we can conclude that our simulated planet is in equilibrium at all modelled pressures. This is quite a significant result, as this allows one to investigate whether a retrieved profile is consistent with 1-D radiative-convective equilibrium throughout the atmosphere, and if not, quantify where in the atmosphere the temperature deviates.

\subsection{Chemical Disequilibrium}

The deviation of the retrieved model atmosphere from chemical equilibrium was calculated by holding the temperature profiles fixed from the retrieval, and determining the equilibrium abundances as described in section \ref{methods_chem_diseqm}. Chemical deviation provides a handle on the mixing at play in the atmosphere, e.g. from winds to vertical mixing, and perhaps even the photo-dissociation for some species. 1000 randomly sampled retrieved values were used, and the theoretical equilibrium mixing fractions calculated as a function of the pressure for 100 layers, evenly spaced in log(P) ranging from 100-10$^{-5}$ bar.

\begin{figure*}
	\includegraphics[width=\textwidth]{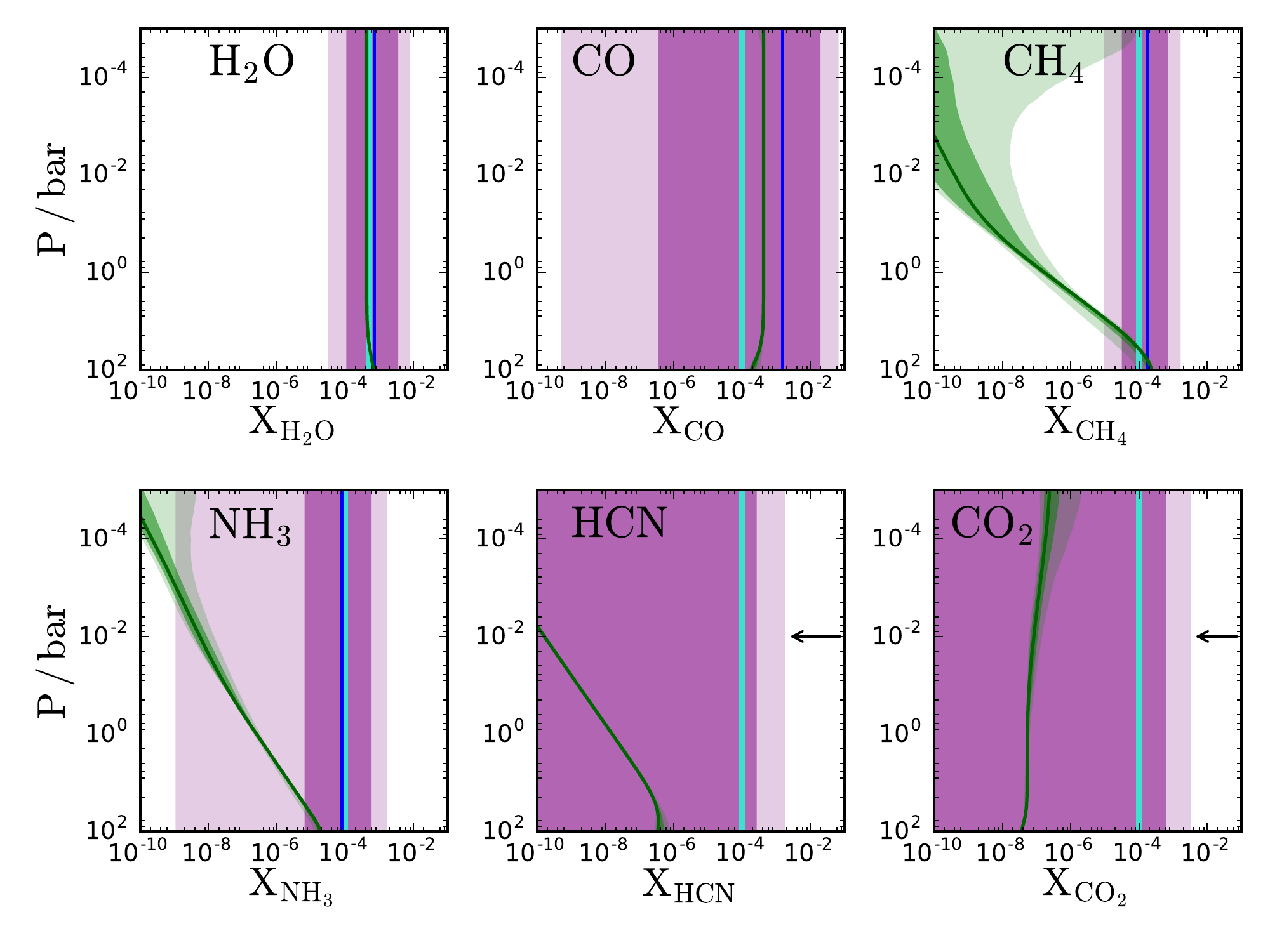}
    \caption{Comparison of retrieved chemical abundances and theoretical chemical equilibrium models for simulated data. The cyan line indicates the actual mixing ratio used to generate the spectrum that was retrieved, the dark and light purple contours show the 1 and 2$\sigma$ errors for the retrieval respectively, and the dark and light green the corresponding equilibrium mixing fractions for each species. Where a molecule was detected in the retrieval, the median fit value is also plotted in blue, and where no abundance could be constrained, the 2$\sigma$ upper bound is shown with an arrow. 1000 randomly sampled retrieval points were used, and 100 layers taken for the model atmosphere. The P-T profile was fixed for each random sample from the posterior.}
    \label{fig:sim_diseqm_chem}
\end{figure*}

Fig. \ref{fig:sim_diseqm_chem} shows the mixing fractions used to generate the spectrum and the retrieved chemical abundances on the simulated data set. We find that all of our retrieved values are consistent with the black line used to generate the model spectrum. The chemical equilibrium values are also shown in fig. \ref{fig:sim_diseqm_chem} for the retrieved P-T profiles, and gives us a handle to the disequilibrium present on a planet. We chose our simulated atmosphere to be out of chemical equilibrium (but in radiative equilibrium) given that equilibrium abundances were too low for many of the species in order for them to be constrained effectively. The higher abundances also highlight any degeneracies that may be present between molecules in the retrieval and demonstrate HyDRA's effectiveness in determining chemical disequilibrium. The predicted CH$_4$, NH$_3$ and HCN abundances decrease with height in the atmosphere, whereas the CO$_2$ mixing fraction remained reasonably constant. It was however at all points less than the H$_2$O abundance, justifying our assumption that the CO$_2$ abundance should not exceed the H$_2$O. We can clearly see from the simulated retrieval that the model planet's spectrum is not in equilibrium.

\section{Results}
\label{results}

We now apply HyDRA to retrieve the dayside atmospheric properties of WASP-43b using observed spectra. \cite{kreidberg_2014} and \cite{stevenson_2014} reported detailed retrieval analyses for WASP-43b and reported constraints on its H$_2$O abundance, $P$-$T$ profile, and day-night energy circulation. In the present work, we use the observed thermal emission spectra from these previous works and revisit the constraints on the atmospheric properties using HyDRA. In addition to previous constraints we specifically investigate the deviations of the derived compositions and $P$-$T$ profile from chemical and radiative-convective equilibrium. We take the radius of the planet to be 0.93 R$_J$, with a semi-major axis of 0.014 A.U. and a log(g) = 3.7 (taken from exoplanets.org).

\begin{figure*}
\begin{overpic}[width=\textwidth]{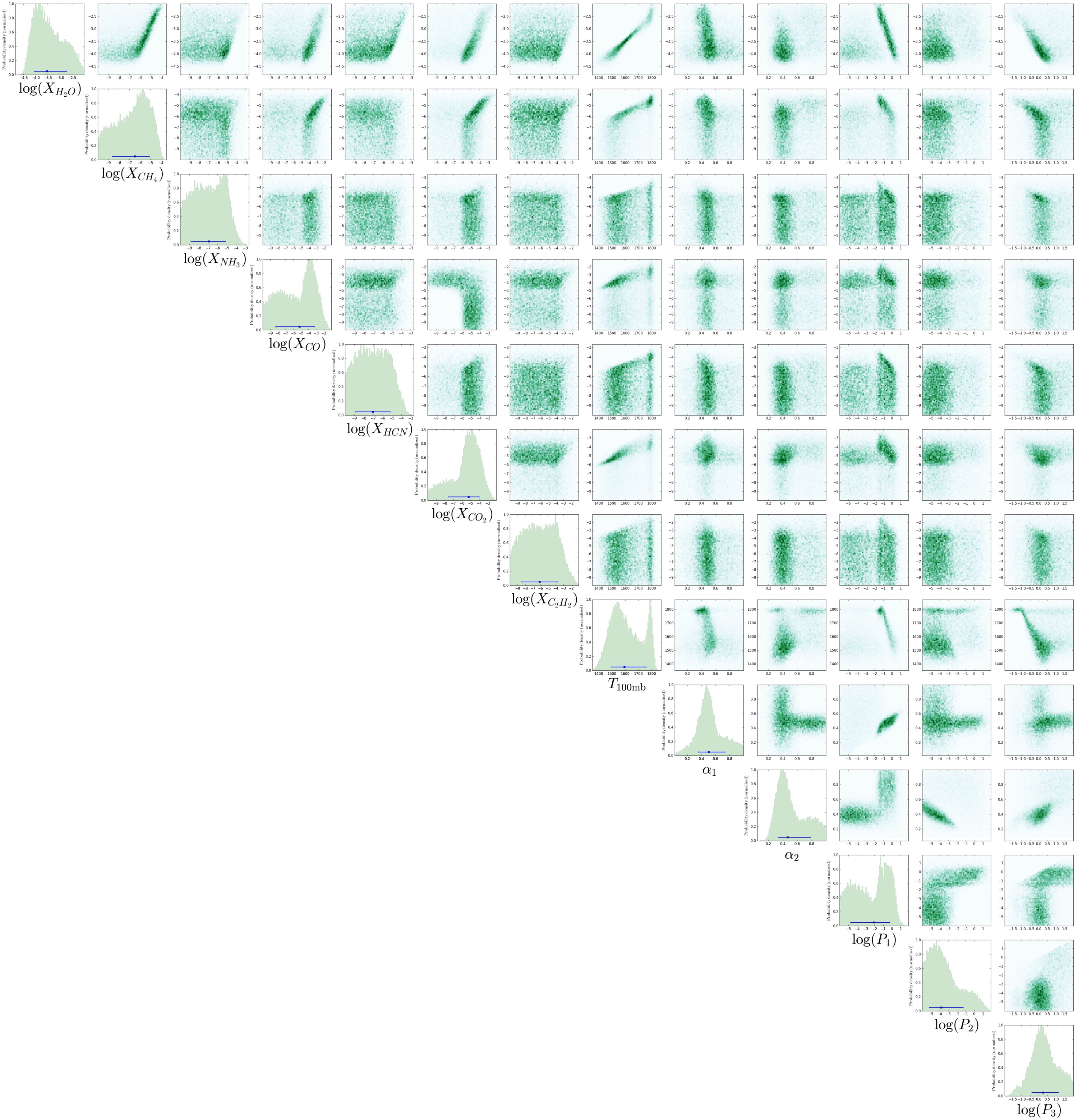}
\put (10,37) {\begin{tabular}{l{c}r}
Parameter              & Value & Error \\
\hline
log(X$_\mathrm{H_2O}$) & -3.54 & $\substack{+0.82 \\ -0.52}$ \\
log(X$_\mathrm{CH_4}$) & < -4.3 & -  \\
log(X$_\mathrm{NH_3}$) & < -4.1 & -  \\
log(X$_\mathrm{CO  }$) & < -2.1 & -  \\
log(X$_\mathrm{HCN }$) & < -4.0 & -  \\
log(X$_\mathrm{CO_2}$) & < -3.0 & -  \\
log(X$_\mathrm{C_2H_2}$) & < -2.3 & -   \\
T$_\mathrm{100mb}$ /K     & 1594 & $\substack{+170 \\ -101}$ \\
$\alpha_1 \, /\mathrm{K}^{-\frac{1}{2}}$     & 0.50 & $\substack{+0.24 \\ -0.15}$ \\
$\alpha_2 \, /\mathrm{K}^{-\frac{1}{2}}$     & 0.47 & $\substack{+0.32 \\ -0.14}$ \\
log(P$_1$/bar)     & -2.1 & $\substack{+1.8 \\ -2.7}$ \\
log(P$_2$/bar)     & -3.8 & $\substack{+2.5 \\ -1.4}$ \\
log(P$_3$/bar)     & 0.26 &$\substack{+0.91 \\ -0.67}$\\
\end{tabular}
}
\end{overpic}
    \caption{Marginalised posterior distribution of WASP-43b's atmosphere under emission spectroscopy. The data set used for the retrieval was obtained from \citet{kreidberg_2014} and considers the Hubble WFC3 and Spitzer 3.6$\micron$ and 4.5$\micron$ channels. We considered 7 molecular volatile species and six parameters describing the P-T profile of the atmosphere. 4000 evenly spaced points in wavelength were used to generate spectra between 1$\micron$ and 5.5 $\micron$, with 4000 live points used for the nested sampling and 100 atmospheric layers, with over $10^6$ models run in total. The histograms and relative correlations between the retrieval parameters is shown on the top right hand side. The table shows the retrieved values and their associated 1$\sigma$ error bars. The upper limits where shown are 2$\sigma$ upper bounds.}
    \label{fig:wasp43_ret_corner}
\end{figure*}

\begin{figure*}
	\includegraphics[width=\textwidth]{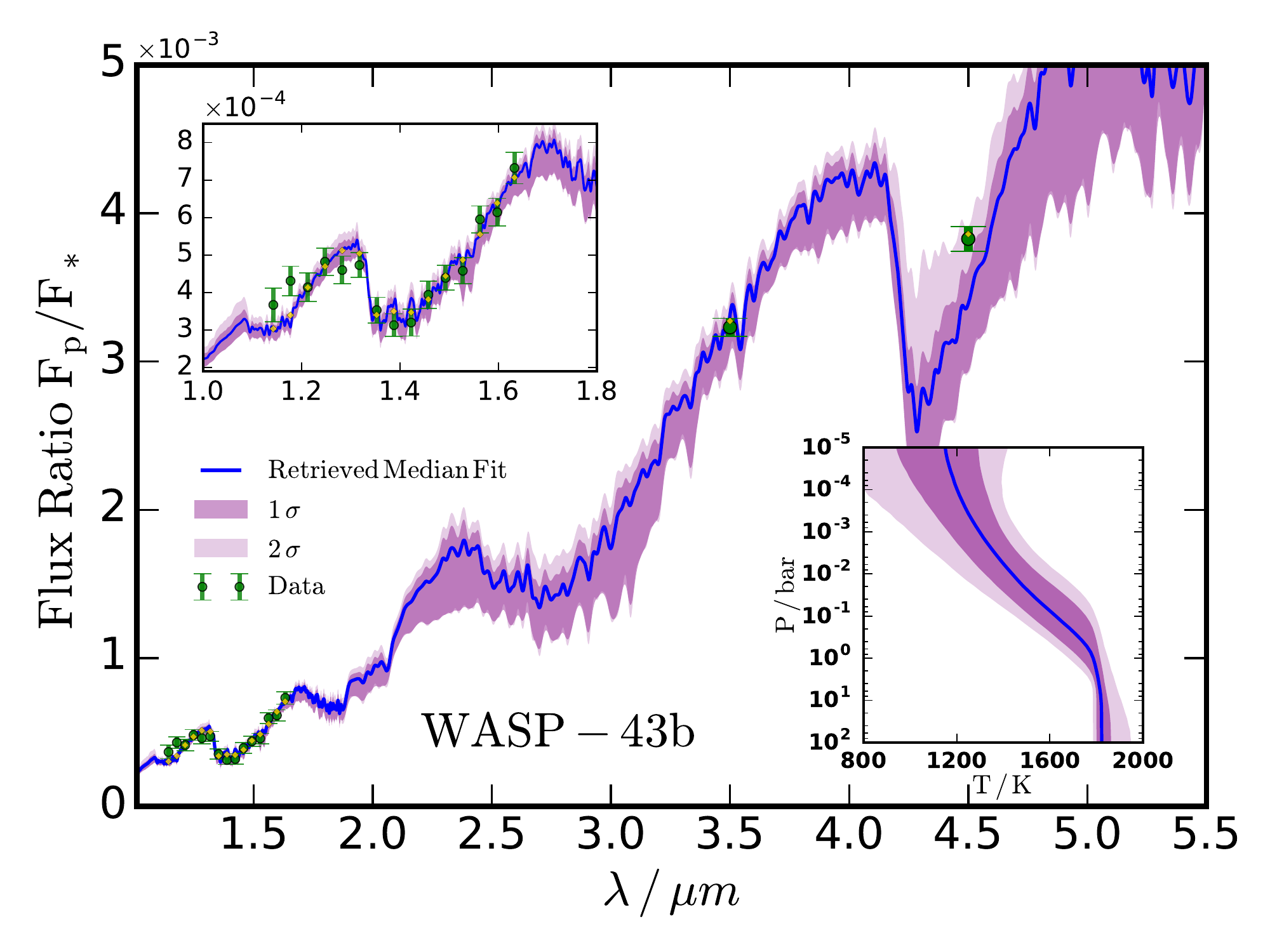}
    \caption{Retrieved emission spectrum of WASP-43b, showing the median fit spectrum and the 1 and 2$\sigma$ uncertainty. The green markers indicate the data set and the corresponding error bars, and the yellow diamonds the binned median model. The top left inset shows the WFC3 bandpass and the corresponding data points and spectral fit. The bottom right inset shows the retrieved P-T profile and the corresponding retrieved error on the temperature.}
    \label{fig:wasp43_spectrum}
\end{figure*}

Following previous works we consider high-precision HST and Spitzer data for the present analysis. The WFC3 data, in the 1.1-1.7 $\mu$m range, were obtained from \cite{kreidberg_2014} and two Spitzer IRAC photometric points, at 3.6 $\micron$ and 4.5 $\micron$ were obtained from \cite{blecic_2014}. The model parameters, as described in section~\ref{methods}, include mixing ratios of the prominent chemical species (H$_2$O, CH$_4$, NH$_3$, CO, HCN, CO$_2$ and C$_2$H$_2$) and the parameters of the $P$-$T$ profile. The retrievals, carried out using the nested sampling algorithm, used 4000 live samples and included over $10^6$ model evaluations per retrieval. The detection significances for the chemical species were carried out using nested model comparisons \citep[see e.g.][]{macdonald_2017}, resulting in $\sim$10$^7$ model evaluations overall. Section \ref{bayes_factor} has further information on the Bayesian model comparisons and describes how we evaluated the detection significance of a species given the evidence $\mathcal{Z}$ for each model with and without the relevant species. The results of the retrieval are shown in figs.~\ref{fig:wasp43_ret_corner} and \ref{fig:wasp43_spectrum} with deviations from equilibrium given in figs.~\ref{fig:diseqm_rc} and \ref{fig:diseqm_chem}. 

\subsection{Constraints on Chemical Abundances}

We report strong constraints on the abundance of H$_2$O which is detected at 8-$\sigma$ significance. The constraints on the chemical abundances of all species considered are shown in fig.~\ref{fig:wasp43_ret_corner}. The volume mixing ratio of H$_2$O is retrieved to be $\log{\rm X_{H_2O}} = -3.54^{+0.82}_{-0.52}$ is consistent with that of \cite{kreidberg_2014} within 1-$\sigma$, and is consistent with expectations from a solar abundance atmosphere at the temperature of WASP-43b. We also note the H$_2$O-T$_\mathrm{100mb}$ degeneracy in fig. \ref{fig:wasp43_ret_corner} as seen for our simulated data and discussed in section \ref{h2o_degeneracy}. The corresponding spectral fit is shown in fig. \ref{fig:wasp43_spectrum}. The strong constraint on the H$_2$O abundance is made possible by the strong H$_2$O features in the WFC3 bandpass. 

We also report a joint detection of CO or CO$_2$ in the atmosphere. The constraint on these two species arise from the strong features of both species in the Spitzer 4.5 $\micron$ bandpass. This also leads to a degeneracy between the two molecules due to a lack of strong features from either molecule elsewhere in the observed spectral range. The degeneracy is apparent in the ``L"-shaped feature in the CO-CO$_2$ correlation plot in fig.\ref{fig:wasp43_ret_corner}. Neither CO nor CO$_2$ have individual significant detection evidences (1.3 and 2.1 $\sigma$, respectively), however we do find that the combined significance CO/CO$_2$ is 7.9$\sigma$. This amounts to a strong detection of carbon chemistry in WASP-43b.

We do not find conclusive evidence for any of the remaining species considered with only upper-limits retrieved. CH$_4$, NH$_3$ and HCN are constrained to be less than $\sim10^{-4}$ at 2-$\sigma$ confidence; they are ostensibly present in smaller quantities than can be retrieved with the current data. As discussed below, equilibrium calculations would also predict low abundances of these species given the temperature. Therefore, their non-detections are perhaps unsurprising. No meaningful constraint is obtained for C$_2$H$_2$, as its cross-section is too weak or has very low abundance and hence is dominated by other species in the WFC3 and Spitzer bandpasses. 

\subsection{Retrieved P-T Profile}

The observed spectrum provides robust constraints on the $P$-$T$ profile of the dayside atmosphere. The retrieved P-T profile along with confidence contours is shown on the inset in fig. \ref{fig:wasp43_spectrum}. The results show the clear absence of a temperature inversion in the observable dayside atmosphere, in agreement with \cite{kreidberg_2014} and \cite{stevenson_2014}. The strong absorption features observed in the spectrum, both in the WFC3 and Spitzer 4.5 $\micron$ bands, constrain the temperature profile to be monotonically decreasing outward in the observable atmosphere. The derived profile also shows an isothermal temperature structure below the photosphere which is characteristic of irradiated hot Jupiters \citep[e.g.][]{burrows_2008,gandhi_2017}. 

The data also provide a strong constraint on the photospheric temperature. The temperature at 100 mb is constrained to be $1594^{+170}_{-101}$ K, which is in nearly exact agreement with \cite{kreidberg_2014}. The derived photospheric temperature is also consistent with the equilibrium temperature without efficient redistribution ($\sim$1635 K), which is also indicated by  thermal phase curve observations \citep{stevenson_2014} as discussed below. The most stringent constraint on the temperature is at $\sim$1 bar, near to where the photosphere ($\tau_\nu \sim 1$) is located. The uncertainty on the temperature increases away from the photosphere on either side and is highest on the two ends of the profile at pressures which are inaccessible to observations as expected. 

\subsection{Radiative-Convective Disequilibrium}

Using HyDRA we are able to constrain the layer-by-layer deviation of the temperature profile from radiative equilibrium. For hot Jupiters, the observable atmosphere in equilibrium is dominated by radiative energy transport. However, non-equilibrium processes such as those caused by atmospheric dynamics (e.g. winds) could drive the atmosphere out of radiative equilibrium. As discussed in section~\ref{methods} one of the primarily capabilities of HyDRA is to constrain the effects of such properties on the layer-by-layer temperature profile. This is achieved by comparing the retrieved distribution of $P$-$T$ profiles with those obtained in radiative-convective equilibrium for the same chemical compositions obtained from the retrieval. As discussed in section~\ref{methods} the self-consistent models are computed using GENESIS code \citep{gandhi_2017} which has been developed in the same framework as the retrieval code. The constraints on the temperature differentials ($\Delta$T) are evaluated based on 1000 randomly sampled points from the posterior distributions of the retrieved $P$-$T$ profiles and their corresponding compositions. 

\begin{figure*}
	\includegraphics[width=\textwidth]{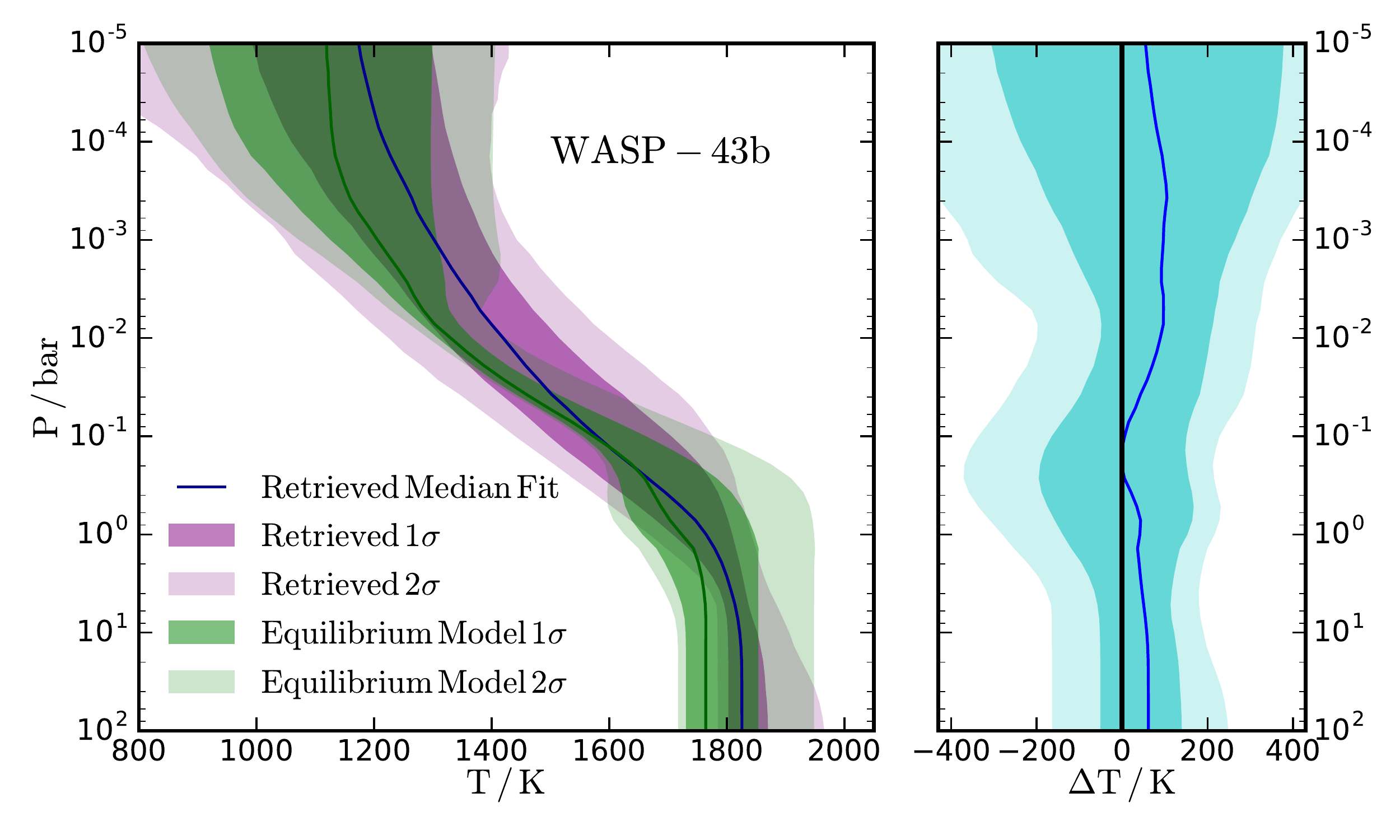}
    \caption{Deviation from 1-D radiative-convective equilibrium present on WASP-43b. The left hand side shows the retrieved P-T profiles in purple and theoretical radiative equilibrium calculations performed with GENESIS \citep{gandhi_2017} in green, with the chemistry fixed to the retrieved values. The right hand side shows the difference between the retrieved and the equilibrium temperatures, with the darker and lighter shade representing the 1 and 2$\sigma$ uncertainties respectively. 1000 randomly sampled points from the retrieval were used to generate the equilibrium models and the sodium and potassium abundances (not retrieved) were set to solar compositions.}
    \label{fig:diseqm_rc}
\end{figure*}

We report the dayside atmosphere of WASP-43b to be in radiative equilibrium with low day-night energy redistribution.  Figure~ \ref{fig:diseqm_rc} shows the constraints on $\Delta$T as a function of pressure and the distributions of both the retrieved and equilibrium $P$-$T$ profiles showing excellent agreement. At all points in the atmosphere the $\Delta$T is consistent to zero within 1-$\sigma$, implying concordance between the retrieved and equilibrium $P$-$T$ profiles. Additionally, the observed agreement is achieved with a low day-night energy redistribution in the equilibrium models. This is also consistent with the inefficient redistribution suggested by \cite{stevenson_2014} based on the large day-night temperature contrast observed with thermal phase curves of the planet. The transition to the isotherm in the lower atmosphere occurs at $\sim1$bar, in agreement with what would be expected in equilibrium. This is the region of the atmosphere where the optical depth exceeds 1 and hence the photons become diffusive, which leads to a profile that does not vary with the pressure. This isothermal structure in the lower atmosphere is consistent with predictions from self-consistent equilibrium models of highly irradiated hot Jupiters \citep{burrows_2008,fortney_2008,gandhi_2017}. As shown in these studies, the transition from the isothermal radiative structure to a convective adiabat occurs much deeper in the atmosphere, at pressures of $\gtrsim$100 bar, which are well below the observable atmosphere. As such, for our equilibrium model we assumed the internal flux of the planet to be negligible as it only effects the deep adiabat which is not accessible to current observations.  

Future observations in other spectral regions will be able to further constrain the temperature profile and deviations from radiative equilibrium. We also assume that the sodium/potassium abundance is solar, and that no other visible absorbers are present in any significant quantities. Visible absorbers such as TiO are likely to be present only in small quantities, as otherwise thermal inversions would occur in the photosphere at $\sim 0.1-0.01$ bar \citep{gandhi_2017}. These models also assume solar composition Na/K, given that the retrieved H$_2$O abundance was consistent with solar. Future compositional estimates on visible absorbers can provide more definitive inputs to the equilibrium models. 

\subsection{Chemical Disequilibrium}

\begin{figure*}
	\includegraphics[width=\textwidth]{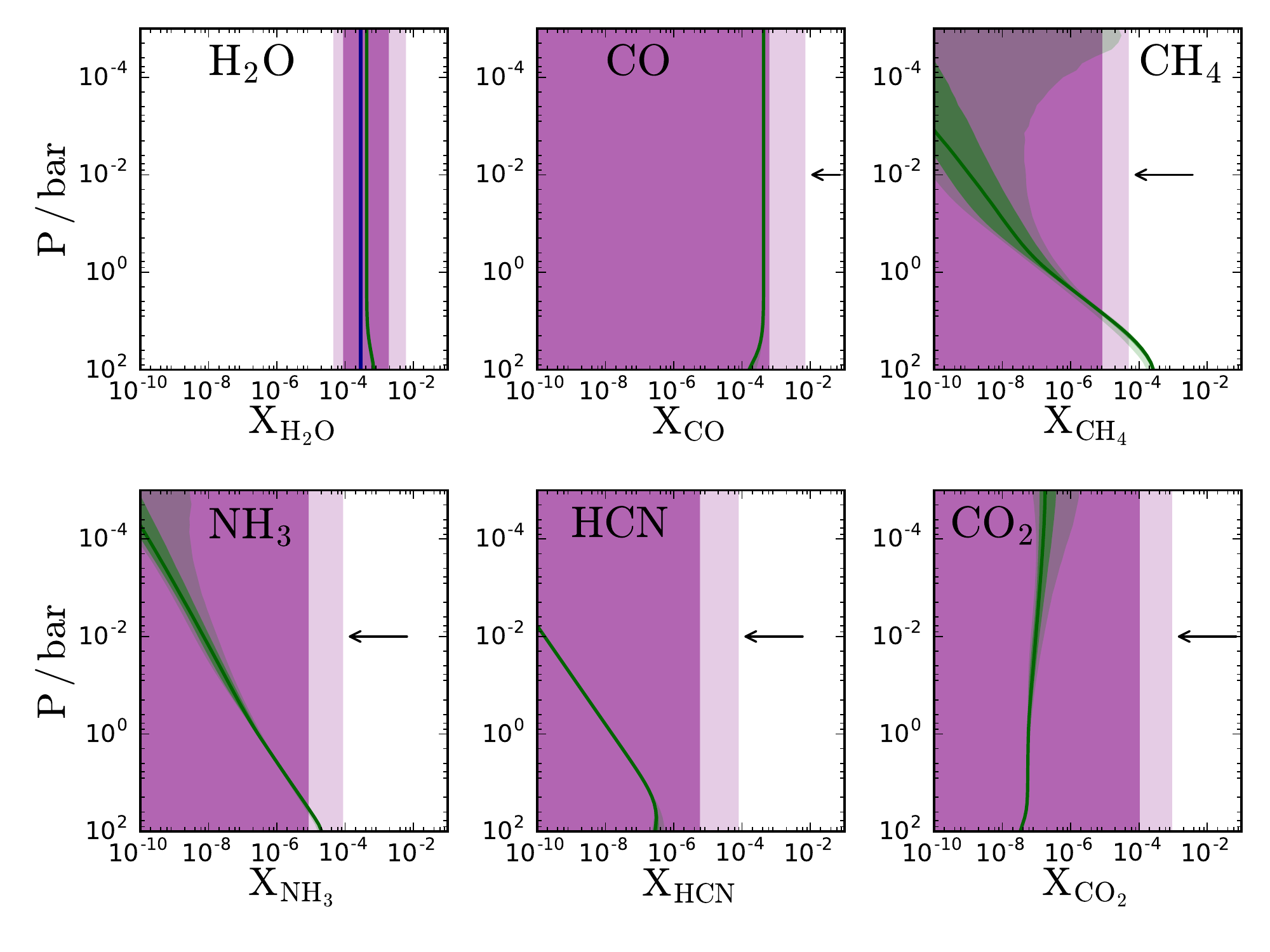}
    \caption{Retrieved chemical abundances and theoretical chemical equilibrium calculations performed with GENESIS \citep{gandhi_2017}. The dark and light purple contours show the 1 and 2$\sigma$ errors for the retrieval respectively, and the dark and light green the corresponding thermochemical equilibrium mixing fractions with the P-T profile fixed to the retrieved values. Where a molecule was detected using our Bayesian analysis, the median fit value is also plotted in blue, and where there was no significant detection, the 2$\sigma$ upper bound is shown by an arrow. 1000 randomly sampled retrieval points were used, and 100 layers taken for the model atmosphere. The atomic abundances were kept at solar values for C, O and N during the equilibrium calculations.}
    \label{fig:diseqm_chem}
\end{figure*}

Here we investigate the possible deviation of the retrieved chemical abundances from thermochemical equilibrium. To do so, we compare the retrieved distributions of the chemical abundances, with their corresponding $P$-$T$ profiles, against those obtained assuming chemical equilibrium for the same $P$-$T$ profiles. The approach is discussed in section~\ref{methods}. The retrieved and equilibrium mixing ratios for the prominent chemical species are shown in fig. \ref{fig:diseqm_chem}. The retrieved H$_2$O abundance is consistent with that obtained in thermochemical equilibrium with the assumption of solar composition. It is important to note that even in chemical equilibrium the H$_2$O abundance is expected to be relatively uniform with depth, as assumed in the retrievals, unlike some other species. This is owing to the fact that at high temperatures, with a solar abundance C/O ratio of 0.5, H$_2$O is expected to be the dominant carrier of oxygen throughout the observable atmosphere \citep{madhu_2012,moses_2013}. 

While only upper-limits are available for molecules besides H$_2$O we nevertheless find the constraints to be consistent with equilibrium expectations. Furthermore, with the exception of H$_2$O and CO, the remaining species are also expected to be present in small quantities in equilibrium. For example, the CH$_4$ abundance in equilibrium decreases outward in the atmosphere with decreasing pressure and is well below the observed upper-limit in the photosphere. In principle, strong vertical mixing could dredge up a higher abundance of CH$_4$ from deeper layers of the atmosphere \citep[e.g.][]{moses_2013}. However, out current upper-limit on CH$_4$ suggests that the corresponding quench pressure level from where such dredge up might occur could not be greater than $\sim$10 bar. Generally, the retrieved CH$_4$ abundance $\lesssim10^{-5}$, is in agreement with our chemical equilibrium predictions. Equilibrium chemistry suggests that the NH$_3$ and HCN should be present in significant amounts only in the deeper layers of the atmosphere, greater than 1 bar in the atmosphere, below which the atmosphere is opaque to radiation ($\tau_\nu>>1$). They can also be dredged up from the vertical mixing into the photosphere \citep{moses_2013,macdonald_2017} though no strong evidence for the same is seen here. 

\section{Discussion and Summary}
\label{discussion}

In this study we introduce HyDRA, a new atmospheric retrieval framework for emission spectroscopy of exoplanets. In addition to the functionalities of existing retrieval codes, HyDRA is geared towards constraining deviations of retrieved solutions from chemical and radiative equilibrium. Emission spectroscopy allows probing the temperature structure and chemical composition of the dayside atmospheres of transiting exoplanets. In the present work, we develop a common framework to operate a state-of-the-art retrieval algorithm in tandem with self-consistent equilibrium models of thermal emission from exoplanets. 

We test HyDRA using simulated data of the hot Jupiter WASP-43b and demonstrate accurate retrieval of the abundances and temperature structure, along with deviations from equilibrium. Consistently with previous work in the literature, we find degeneracies between the retrieved abundances of carbon monoxide and carbon dioxide given the limited observations currently available. This degeneracy was partially broken by imposing the constraint that the H$_2$O abundance must exceed the CO$_2$ abundance, as this is always the case for H$_2$-rich atmospheres \citep{moses_2011,heng_lyons_2016}, but the retrieved abundance is still degenerate with CO. We also see a degeneracy between abundances and the photosphere temperature, where we parametrise the temperature as $T_\mathrm{100mb}$. This degeneracy may be broken with sufficient spectral coverage, particularly data in regions where the continuum opacity is probed (e.g 2-2.5$\micron$). With the future development of instruments that do so, this should be resolved to give us the tightest constraints on the chemistry present on exoplanets.

Whilst previous work has already been done in this field, our model combines several modules from our 1-D self-consistent forward model, GENESIS \citep{gandhi_2017}, such as a similar radiative transfer scheme (which has been tested against the self-consistent solution in section \ref{methods_rad_transfer}) and identical molecular cross-section calculations. The Kurucz stellar model and the line lists for the opacity calculations are also shared. All of this means that we are able to compare quantitatively the difference between retrieved and equilibrium models for a given data set of a planet, and constrain the chemical and radiative-convective disequilibrium. The latter in particular is novel and allows for the precise determination of the deviation in the temperature profile from what is expected for a 1-D model. This opens up many avenues to expand upon our current understanding, and improve our modelling techniques.

We demonstrate our retrieval algorithm on the emission spectrum of WASP-43b, with one of the most precise observations of exoplanets under emission to date \citep{kreidberg_2014}, as a demonstration of our code. We find that the temperature structure of WASP-43b shows agreement with our radiative-convective and chemical equilibrium model. By taking the chemistry to be that which was retrieved, we are able to conclude that this planet's temperature profile is consistent with our forward model. In running our model we assume there is little stellar flux redistribution over to the nightside, and find results in agreement with \cite{stevenson_2017}.

The volatiles' abundances obtained reveals that the planet is most likely close to solar composition water and provides some constraints on the CO abundance. The C/O ratio is likely to be less than 1, particularly given the high water abundance observed and our chemical equilibrium calculations in section \ref{results}. The other species have not been detected to any high significance. The P-T profile indicates that there is no thermal inversion present in the region of $\tau\sim1$. The 100mb temperature is constrained to be $1594\substack{+170 \\ -101}$ K, in agreement with the result from \cite{kreidberg_2014}. The chemical equilibrium analysis indicates that the planet is within chemical equilibrium given the retrieved temperature, albeit with large uncertainties on the results, as lower abundances do not significantly affect the spectrum and hence cannot be retrieved. 

There has been some suggestions \citep{hubeny_2003,fortney_2008} that thermal inversions may be present on hot Jupiters, thanks to strong visible absorption from metallic species such as TiO and VO. We find no such inversions for WASP-43b, as it is most likely too cool for such species to exist in gaseous form. Investigating the spread of thermal inversions on such systems would indirectly tell us about the visible absorption present, and the implications this would mean about their temperature structure.

Work has already begun looking into phase resolved spectroscopy \citep{stevenson_2017}, combining dayside and nightside retrievals dependant on the observational geometry. Modification of the retrieval to take into account the likely temperature distribution would be prudent, as it may help constrain the abundances and the profile better, and provide evidence for temperature variations across the dayside. The temperature structure of the planet is also sensitive in emission spectroscopy, so a good understanding here is required to get precise planetary properties.

HyDRA is a new emission spectroscopy retrieval algorithm that utilises the most up to date high temperature molecular data along with the most advanced radiative transfer techniques and the latest statistical methods. There is much excitement over the future of the field with new and upcoming observations with VLT, JWST, E-ELT and many others, and the present work prepares us for this new era of atmospheric observations. This will allow for even deeper insights into the structure and composition of exoplanetary atmospheres. HyDRA used for emission spectroscopy retrieval in conjunction with our self-consistent GENESIS model \citep{gandhi_2017} represents a new development in the field and allows constraints on departures from 1-D equilibrium in exoplanetary atmospheres, a vital step towards detailed characterization of atmospheric processes in exoplanets. 

\section*{Acknowledgements}

SG thanks Arazi Pinhas for helpful discussions on the code for the parametric model, and Ryan MacDonald for discussions on retrieval methodology. SG acknowledges financial support from the Science and Technology Facilities Council (STFC), UK, towards his doctoral programme.




\bibliographystyle{mnras}
\bibliography{references} 

\begin{thebibliography}{}
\makeatletter
\relax
\def\mn@urlcharsother{\let\do\@makeother \do\$\do\&\do\#\do\^\do\_\do\%\do\~}
\def\mn@doi{\begingroup\mn@urlcharsother \@ifnextchar [ {\mn@doi@}
  {\mn@doi@[]}}
\def\mn@doi@[#1]#2{\def\@tempa{#1}\ifx\@tempa\@empty \href
  {http://dx.doi.org/#2} {doi:#2}\else \href {http://dx.doi.org/#2} {#1}\fi
  \endgroup}
\def\mn@eprint#1#2{\mn@eprint@#1:#2::\@nil}
\def\mn@eprint@arXiv#1{\href {http://arxiv.org/abs/#1} {{\tt arXiv:#1}}}
\def\mn@eprint@dblp#1{\href {http://dblp.uni-trier.de/rec/bibtex/#1.xml}
  {dblp:#1}}
\def\mn@eprint@#1:#2:#3:#4\@nil{\def\@tempa {#1}\def\@tempb {#2}\def\@tempc
  {#3}\ifx \@tempc \@empty \let \@tempc \@tempb \let \@tempb \@tempa \fi \ifx
  \@tempb \@empty \def\@tempb {arXiv}\fi \@ifundefined
  {mn@eprint@\@tempb}{\@tempb:\@tempc}{\expandafter \expandafter \csname
  mn@eprint@\@tempb\endcsname \expandafter{\@tempc}}}

\bibitem[\protect\citeauthoryear{{Benneke} \& {Seager}}{{Benneke} \&
  {Seager}}{2013}]{benneke_2013}
{Benneke} B.,  {Seager} S.,  2013, \mn@doi [\apj]
  {10.1088/0004-637X/778/2/153}, \href
  {http://adsabs.harvard.edu/abs/2013ApJ...778..153B} {778, 153}

\bibitem[\protect\citeauthoryear{{Birkby}, {de Kok}, {Brogi}, {de Mooij},
  {Schwarz}, {Albrecht}  \& {Snellen}}{{Birkby} et~al.}{2013}]{birkby_2013}
{Birkby} J.~L.,  {de Kok} R.~J.,  {Brogi} M.,  {de Mooij} E.~J.~W.,  {Schwarz}
  H.,  {Albrecht} S.,   {Snellen} I.~A.~G.,  2013, \mn@doi [\mnras]
  {10.1093/mnrasl/slt107}, \href
  {http://adsabs.harvard.edu/abs/2013MNRAS.436L..35B} {436, L35}

\bibitem[\protect\citeauthoryear{{Blecic} et~al.,}{{Blecic}
  et~al.}{2014}]{blecic_2014}
{Blecic} J.,  et~al., 2014, \mn@doi [\apj] {10.1088/0004-637X/781/2/116}, \href
  {http://adsabs.harvard.edu/abs/2014ApJ...781..116B} {781, 116}

\bibitem[\protect\citeauthoryear{{Buchner} et~al.,}{{Buchner}
  et~al.}{2014}]{buchner_2014}
{Buchner} J.,  et~al., 2014, \mn@doi [\aap] {10.1051/0004-6361/201322971},
  \href {http://adsabs.harvard.edu/abs/2014A%26A...564A.125B} {564, A125}

\bibitem[\protect\citeauthoryear{{Burningham}, {Marley}, {Line}, {Lupu},
  {Visscher}, {Morley}, {Saumon}  \& {Freedman}}{{Burningham}
  et~al.}{2017}]{burningham_2017}
{Burningham} B.,  {Marley} M.~S.,  {Line} M.~R.,  {Lupu} R.,  {Visscher} C.,
  {Morley} C.~V.,  {Saumon} D.,   {Freedman} R.,  2017, \mn@doi [\mnras]
  {10.1093/mnras/stx1246}, \href
  {http://adsabs.harvard.edu/abs/2017MNRAS.470.1177B} {470, 1177}

\bibitem[\protect\citeauthoryear{{Burrows}, {Budaj}  \& {Hubeny}}{{Burrows}
  et~al.}{2008}]{burrows_2008}
{Burrows} A.,  {Budaj} J.,   {Hubeny} I.,  2008, \mn@doi [\apj]
  {10.1086/533518}, \href {http://adsabs.harvard.edu/abs/2008ApJ...678.1436B}
  {678, 1436}

\bibitem[\protect\citeauthoryear{{Castelli} \& {Kurucz}}{{Castelli} \&
  {Kurucz}}{2004}]{kurucz_model}
{Castelli} F.,  {Kurucz} R.~L.,  2004, preprint, \href
  {http://adsabs.harvard.edu/abs/2004astro.ph..5087C} {} (\mn@eprint {}
  {astro-ph/0405087})

\bibitem[\protect\citeauthoryear{{Deming} et~al.,}{{Deming}
  et~al.}{2013}]{deming_2013}
{Deming} D.,  et~al., 2013, \mn@doi [\apj] {10.1088/0004-637X/774/2/95}, \href
  {http://adsabs.harvard.edu/abs/2013ApJ...774...95D} {774, 95}

\bibitem[\protect\citeauthoryear{{Diamond-Lowe}, {Stevenson}, {Bean}, {Line}
  \& {Fortney}}{{Diamond-Lowe} et~al.}{2014}]{diamond-lowe_2014}
{Diamond-Lowe} H.,  {Stevenson} K.~B.,  {Bean} J.~L.,  {Line} M.~R.,
  {Fortney} J.~J.,  2014, \mn@doi [\apj] {10.1088/0004-637X/796/1/66}, \href
  {http://adsabs.harvard.edu/abs/2014ApJ...796...66D} {796, 66}

\bibitem[\protect\citeauthoryear{{Evans} et~al.,}{{Evans}
  et~al.}{2017}]{evans_2017}
{Evans} T.~M.,  et~al., 2017, \mn@doi [\nat] {10.1038/nature23266}, \href
  {http://adsabs.harvard.edu/abs/2017Natur.548...58E} {548, 58}

\bibitem[\protect\citeauthoryear{{Feroz} \& {Hobson}}{{Feroz} \&
  {Hobson}}{2008}]{feroz_2008}
{Feroz} F.,  {Hobson} M.~P.,  2008, \mn@doi [\mnras]
  {10.1111/j.1365-2966.2007.12353.x}, \href
  {http://adsabs.harvard.edu/abs/2008MNRAS.384..449F} {384, 449}

\bibitem[\protect\citeauthoryear{{Feroz}, {Hobson}  \& {Bridges}}{{Feroz}
  et~al.}{2009}]{feroz_2009}
{Feroz} F.,  {Hobson} M.~P.,   {Bridges} M.,  2009, \mn@doi [\mnras]
  {10.1111/j.1365-2966.2009.14548.x}, \href
  {http://adsabs.harvard.edu/abs/2009MNRAS.398.1601F} {398, 1601}

\bibitem[\protect\citeauthoryear{{Feroz}, {Hobson}, {Cameron}  \&
  {Pettitt}}{{Feroz} et~al.}{2013}]{feroz_2013}
{Feroz} F.,  {Hobson} M.~P.,  {Cameron} E.,   {Pettitt} A.~N.,  2013, preprint,
  \href {http://adsabs.harvard.edu/abs/2013arXiv1306.2144F} {} (\mn@eprint
  {arXiv} {1306.2144})

\bibitem[\protect\citeauthoryear{{Fortney}, {Lodders}, {Marley}  \&
  {Freedman}}{{Fortney} et~al.}{2008}]{fortney_2008}
{Fortney} J.~J.,  {Lodders} K.,  {Marley} M.~S.,   {Freedman} R.~S.,  2008,
  \mn@doi [APJ] {10.1086/528370}, \href
  {http://adsabs.harvard.edu/abs/2008ApJ...678.1419F} {678, 1419}

\bibitem[\protect\citeauthoryear{{Gandhi} \& {Madhusudhan}}{{Gandhi} \&
  {Madhusudhan}}{2017}]{gandhi_2017}
{Gandhi} S.,  {Madhusudhan} N.,  2017, \mn@doi [\mnras]
  {10.1093/mnras/stx1601}, \href
  {http://adsabs.harvard.edu/abs/2017MNRAS.472.2334G} {472, 2334}

\bibitem[\protect\citeauthoryear{{Guillot}}{{Guillot}}{2010}]{guillot_2010}
{Guillot} T.,  2010, \mn@doi [\aap] {10.1051/0004-6361/200913396}, \href
  {http://adsabs.harvard.edu/abs/2010A%26A...520A..27G} {520, A27}

\bibitem[\protect\citeauthoryear{{Haynes}, {Mandell}, {Madhusudhan}, {Deming}
  \& {Knutson}}{{Haynes} et~al.}{2015}]{haynes_2015}
{Haynes} K.,  {Mandell} A.~M.,  {Madhusudhan} N.,  {Deming} D.,   {Knutson} H.,
   2015, \mn@doi [\apj] {10.1088/0004-637X/806/2/146}, \href
  {http://adsabs.harvard.edu/abs/2015ApJ...806..146H} {806, 146}

\bibitem[\protect\citeauthoryear{{Hellier} et~al.,}{{Hellier}
  et~al.}{2011}]{hellier_2011}
{Hellier} C.,  et~al., 2011, \mn@doi [\aap] {10.1051/0004-6361/201117081},
  \href {http://adsabs.harvard.edu/abs/2011A%26A...535L...7H} {535, L7}

\bibitem[\protect\citeauthoryear{{Heng} \& {Lyons}}{{Heng} \&
  {Lyons}}{2016}]{heng_lyons_2016}
{Heng} K.,  {Lyons} J.~R.,  2016, \mn@doi [\apj] {10.3847/0004-637X/817/2/149},
  \href {http://adsabs.harvard.edu/abs/2016ApJ...817..149H} {817, 149}

\bibitem[\protect\citeauthoryear{{Heng} \& {Tsai}}{{Heng} \&
  {Tsai}}{2016}]{heng_2016}
{Heng} K.,  {Tsai} S.-M.,  2016, \mn@doi [\apj] {10.3847/0004-637X/829/2/104},
  \href {http://adsabs.harvard.edu/abs/2016ApJ...829..104H} {829, 104}

\bibitem[\protect\citeauthoryear{{Hubeny}, {Burrows}  \& {Sudarsky}}{{Hubeny}
  et~al.}{2003}]{hubeny_2003}
{Hubeny} I.,  {Burrows} A.,   {Sudarsky} D.,  2003, \mn@doi [\apj]
  {10.1086/377080}, \href {http://adsabs.harvard.edu/abs/2003ApJ...594.1011H}
  {594, 1011}

\bibitem[\protect\citeauthoryear{{Kataria}, {Showman}, {Fortney}, {Stevenson},
  {Line}, {Kreidberg}, {Bean}  \& {D{\'e}sert}}{{Kataria}
  et~al.}{2015}]{kataria_2015}
{Kataria} T.,  {Showman} A.~P.,  {Fortney} J.~J.,  {Stevenson} K.~B.,  {Line}
  M.~R.,  {Kreidberg} L.,  {Bean} J.~L.,   {D{\'e}sert} J.-M.,  2015, \mn@doi
  [\apj] {10.1088/0004-637X/801/2/86}, \href
  {http://adsabs.harvard.edu/abs/2015ApJ...801...86K} {801, 86}

\bibitem[\protect\citeauthoryear{{Konopacky}, {Barman}, {Macintosh}  \&
  {Marois}}{{Konopacky} et~al.}{2013}]{konopacky_2013}
{Konopacky} Q.~M.,  {Barman} T.~S.,  {Macintosh} B.~A.,   {Marois} C.,  2013,
  \mn@doi [Science] {10.1126/science.1232003}, \href
  {http://adsabs.harvard.edu/abs/2013Sci...339.1398K} {339, 1398}

\bibitem[\protect\citeauthoryear{{Kreidberg} et~al.,}{{Kreidberg}
  et~al.}{2014}]{kreidberg_2014}
{Kreidberg} L.,  et~al., 2014, \mn@doi [\apjl] {10.1088/2041-8205/793/2/L27},
  \href {http://adsabs.harvard.edu/abs/2014ApJ...793L..27K} {793, L27}

\bibitem[\protect\citeauthoryear{{Kurucz}}{{Kurucz}}{1979}]{Kurucz_1979_paper}
{Kurucz} R.~L.,  1979, \mn@doi [\apjs] {10.1086/190589}, \href
  {http://adsabs.harvard.edu/abs/1979ApJS...40....1K} {40, 1}

\bibitem[\protect\citeauthoryear{{Lavie} et~al.,}{{Lavie}
  et~al.}{2017}]{lavie_2017}
{Lavie} B.,  et~al., 2017, \mn@doi [\aj] {10.3847/1538-3881/aa7ed8}, \href
  {http://adsabs.harvard.edu/abs/2017AJ....154...91L} {154, 91}

\bibitem[\protect\citeauthoryear{{Lee}, {Fletcher}  \& {Irwin}}{{Lee}
  et~al.}{2012}]{lee_2012}
{Lee} J.-M.,  {Fletcher} L.~N.,   {Irwin} P.~G.~J.,  2012, \mn@doi [\mnras]
  {10.1111/j.1365-2966.2011.20013.x}, \href
  {http://adsabs.harvard.edu/abs/2012MNRAS.420..170L} {420, 170}

\bibitem[\protect\citeauthoryear{{Line} \& {Parmentier}}{{Line} \&
  {Parmentier}}{2016}]{line2_2016}
{Line} M.~R.,  {Parmentier} V.,  2016, \mn@doi [\apj]
  {10.3847/0004-637X/820/1/78}, \href
  {http://adsabs.harvard.edu/abs/2016ApJ...820...78L} {820, 78}

\bibitem[\protect\citeauthoryear{{Line}, {Zhang}, {Vasisht}, {Natraj}, {Chen}
  \& {Yung}}{{Line} et~al.}{2012}]{line_2012}
{Line} M.~R.,  {Zhang} X.,  {Vasisht} G.,  {Natraj} V.,  {Chen} P.,   {Yung}
  Y.~L.,  2012, \mn@doi [\apj] {10.1088/0004-637X/749/1/93}, \href
  {http://adsabs.harvard.edu/abs/2012ApJ...749...93L} {749, 93}

\bibitem[\protect\citeauthoryear{{Line} et~al.,}{{Line}
  et~al.}{2013}]{line_2013}
{Line} M.~R.,  et~al., 2013, \mn@doi [\apj] {10.1088/0004-637X/775/2/137},
  \href {http://adsabs.harvard.edu/abs/2013ApJ...775..137L} {775, 137}

\bibitem[\protect\citeauthoryear{{Line}, {Knutson}, {Wolf}  \& {Yung}}{{Line}
  et~al.}{2014}]{line_2014}
{Line} M.~R.,  {Knutson} H.,  {Wolf} A.~S.,   {Yung} Y.~L.,  2014, \mn@doi
  [\apj] {10.1088/0004-637X/783/2/70}, \href
  {http://adsabs.harvard.edu/abs/2014ApJ...783...70L} {783, 70}

\bibitem[\protect\citeauthoryear{{Line} et~al.,}{{Line}
  et~al.}{2016}]{line_2016}
{Line} M.~R.,  et~al., 2016, \mn@doi [\aj] {10.3847/0004-6256/152/6/203}, \href
  {http://adsabs.harvard.edu/abs/2016AJ....152..203L} {152, 203}

\bibitem[\protect\citeauthoryear{{MacDonald} \& {Madhusudhan}}{{MacDonald} \&
  {Madhusudhan}}{2017}]{macdonald_2017}
{MacDonald} R.~J.,  {Madhusudhan} N.,  2017, \mn@doi [\mnras]
  {10.1093/mnras/stx804}, \href
  {http://adsabs.harvard.edu/abs/2017MNRAS.469.1979M} {469, 1979}

\bibitem[\protect\citeauthoryear{{Macintosh} et~al.,}{{Macintosh}
  et~al.}{2015}]{macintosh_2015}
{Macintosh} B.,  et~al., 2015, \mn@doi [Science] {10.1126/science.aac5891},
  \href {http://adsabs.harvard.edu/abs/2015Sci...350...64M} {350, 64}

\bibitem[\protect\citeauthoryear{{Madhusudhan}}{{Madhusudhan}}{2012}]{madhu_20%
12}
{Madhusudhan} N.,  2012, \mn@doi [APJ] {10.1088/0004-637X/758/1/36}, \href
  {http://adsabs.harvard.edu/abs/2012ApJ...758...36M} {758, 36}

\bibitem[\protect\citeauthoryear{{Madhusudhan} \& {Seager}}{{Madhusudhan} \&
  {Seager}}{2009}]{madhu_2009}
{Madhusudhan} N.,  {Seager} S.,  2009, \mn@doi [\apj]
  {10.1088/0004-637X/707/1/24}, \href
  {http://adsabs.harvard.edu/abs/2009ApJ...707...24M} {707, 24}

\bibitem[\protect\citeauthoryear{{Madhusudhan} \& {Seager}}{{Madhusudhan} \&
  {Seager}}{2011}]{madhu_2011}
{Madhusudhan} N.,  {Seager} S.,  2011, \mn@doi [\apj]
  {10.1088/0004-637X/729/1/41}, \href
  {http://adsabs.harvard.edu/abs/2011ApJ...729...41M} {729, 41}

\bibitem[\protect\citeauthoryear{{Madhusudhan} et~al.,}{{Madhusudhan}
  et~al.}{2011}]{madhu_2011_inv}
{Madhusudhan} N.,  et~al., 2011, \mn@doi [\nat] {10.1038/nature09602}, \href
  {http://adsabs.harvard.edu/abs/2011Natur.469...64M} {469, 64}

\bibitem[\protect\citeauthoryear{Madhusudhan, Crouzet, McCullough, Deming  \&
  Hedges}{Madhusudhan et~al.}{2014a}]{madhu_2014}
Madhusudhan N.,  Crouzet N.,  McCullough P.~R.,  Deming D.,   Hedges C.,
  2014a, The Astrophysical Journal Letters, 791, L9

\bibitem[\protect\citeauthoryear{{Madhusudhan}, {Amin}  \&
  {Kennedy}}{{Madhusudhan} et~al.}{2014b}]{madhu_2014c}
{Madhusudhan} N.,  {Amin} M.~A.,   {Kennedy} G.~M.,  2014b, \mn@doi [\apjl]
  {10.1088/2041-8205/794/1/L12}, \href
  {http://adsabs.harvard.edu/abs/2014ApJ...794L..12M} {794, L12}

\bibitem[\protect\citeauthoryear{{Madhusudhan}, {Apai}  \&
  {Gandhi}}{{Madhusudhan} et~al.}{2016}]{madhu_2016}
{Madhusudhan} N.,  {Apai} D.,   {Gandhi} S.,  2016, preprint, \href
  {http://adsabs.harvard.edu/abs/2016arXiv161203174M} {} (\mn@eprint {arXiv}
  {1612.03174})

\bibitem[\protect\citeauthoryear{{Malik} et~al.,}{{Malik}
  et~al.}{2017}]{malik_2017}
{Malik} M.,  et~al., 2017, \mn@doi [\aj] {10.3847/1538-3881/153/2/56}, \href
  {http://adsabs.harvard.edu/abs/2017AJ....153...56M} {153, 56}

\bibitem[\protect\citeauthoryear{Molli\`{e}re, van Boekel, Dullemond, Henning
  \& Mordasini}{Molli\`{e}re et~al.}{2015}]{molliere_2015}
Molli\`{e}re P.,  van Boekel R.,  Dullemond C.,  Henning T.,   Mordasini C.,
  2015, The Astrophysical Journal, 813, 47

\bibitem[\protect\citeauthoryear{{Moses} et~al.,}{{Moses}
  et~al.}{2011}]{moses_2011}
{Moses} J.~I.,  et~al., 2011, \mn@doi [\apj] {10.1088/0004-637X/737/1/15},
  \href {http://adsabs.harvard.edu/abs/2011ApJ...737...15M} {737, 15}

\bibitem[\protect\citeauthoryear{{Moses}, {Madhusudhan}, {Visscher}  \&
  {Freedman}}{{Moses} et~al.}{2013}]{moses_2013}
{Moses} J.~I.,  {Madhusudhan} N.,  {Visscher} C.,   {Freedman} R.~S.,  2013,
  \mn@doi [\apj] {10.1088/0004-637X/763/1/25}, \href
  {http://adsabs.harvard.edu/abs/2013ApJ...763...25M} {763, 25}

\bibitem[\protect\citeauthoryear{{Oreshenko} et~al.,}{{Oreshenko}
  et~al.}{2017}]{oreshenko_2017}
{Oreshenko} M.,  et~al., 2017, preprint, \href
  {http://adsabs.harvard.edu/abs/2017arXiv170900338O} {} (\mn@eprint {arXiv}
  {1709.00338})

\bibitem[\protect\citeauthoryear{{Richard} et~al.,}{{Richard}
  et~al.}{2012}]{richard_2012}
{Richard} C.,  et~al., 2012, \mn@doi [\jqsrt] {10.1016/j.jqsrt.2011.11.004},
  \href {http://adsabs.harvard.edu/abs/2012JQSRT.113.1276R} {113, 1276}

\bibitem[\protect\citeauthoryear{{Rothman} et~al.,}{{Rothman}
  et~al.}{2010}]{rothman_2010}
{Rothman} L.~S.,  et~al., 2010, \mn@doi [JQSRT] {10.1016/j.jqsrt.2010.05.001},
  \href {http://adsabs.harvard.edu/abs/2010JQSRT.111.2139R} {111, 2139}

\bibitem[\protect\citeauthoryear{{Rothman} et~al.,}{{Rothman}
  et~al.}{2013}]{rothman_2013}
{Rothman} L.~S.,  et~al., 2013, \mn@doi [JQSRT] {10.1016/j.jqsrt.2013.07.002},
  \href {http://adsabs.harvard.edu/abs/2013JQSRT.130....4R} {130, 4}

\bibitem[\protect\citeauthoryear{{Seager}}{{Seager}}{2010}]{seager_2010}
{Seager} S.,  2010, {Exoplanet Atmospheres: Physical Processes}

\bibitem[\protect\citeauthoryear{{Seager}, {Richardson}, {Hansen}, {Menou},
  {Cho}  \& {Deming}}{{Seager} et~al.}{2005}]{seager_2005}
{Seager} S.,  {Richardson} L.~J.,  {Hansen} B.~M.~S.,  {Menou} K.,  {Cho}
  J.~Y.-K.,   {Deming} D.,  2005, \mn@doi [APJ] {10.1086/444411}, \href
  {http://adsabs.harvard.edu/abs/2005ApJ...632.1122S} {632, 1122}

\bibitem[\protect\citeauthoryear{Showman, Fortney, Lian, Marley, Freedman,
  Knutson  \& Charbonneau}{Showman et~al.}{2009}]{showman_2009}
Showman A.~P.,  Fortney J.~J.,  Lian Y.,  Marley M.~S.,  Freedman R.~S.,
  Knutson H.~A.,   Charbonneau D.,  2009, The Astrophysical Journal, 699, 564

\bibitem[\protect\citeauthoryear{{Sing} et~al.,}{{Sing}
  et~al.}{2016}]{sing_2016}
{Sing} D.~K.,  et~al., 2016, \mn@doi [\nat] {10.1038/nature16068}, \href
  {http://adsabs.harvard.edu/abs/2016Natur.529...59S} {529, 59}

\bibitem[\protect\citeauthoryear{{Skilling}}{{Skilling}}{2004}]{skilling_2004}
{Skilling} J.,  2004, in {Fischer} R.,  {Preuss} R.,   {Toussaint} U.~V.,  eds,
   American Institute of Physics Conference Series Vol. 735, American Institute
  of Physics Conference Series. pp 395--405, \mn@doi{10.1063/1.1835238}

\bibitem[\protect\citeauthoryear{{Snellen}, {de Kok}, {de Mooij}  \&
  {Albrecht}}{{Snellen} et~al.}{2010}]{snellen_2010}
{Snellen} I.~A.~G.,  {de Kok} R.~J.,  {de Mooij} E.~J.~W.,   {Albrecht} S.,
  2010, \mn@doi [\nat] {10.1038/nature09111}, \href
  {http://adsabs.harvard.edu/abs/2010Natur.465.1049S} {465, 1049}

\bibitem[\protect\citeauthoryear{{Stevenson} et~al.,}{{Stevenson}
  et~al.}{2010}]{stevenson_2010}
{Stevenson} K.~B.,  et~al., 2010, \mn@doi [\nat] {10.1038/nature09013}, \href
  {http://adsabs.harvard.edu/abs/2010Natur.464.1161S} {464, 1161}

\bibitem[\protect\citeauthoryear{{Stevenson}, {Bean}, {Madhusudhan}  \&
  {Harrington}}{{Stevenson} et~al.}{2014}]{stevenson_2014}
{Stevenson} K.~B.,  {Bean} J.~L.,  {Madhusudhan} N.,   {Harrington} J.,  2014,
  \mn@doi [\apj] {10.1088/0004-637X/791/1/36}, \href
  {http://adsabs.harvard.edu/abs/2014ApJ...791...36S} {791, 36}

\bibitem[\protect\citeauthoryear{{Stevenson} et~al.,}{{Stevenson}
  et~al.}{2017}]{stevenson_2017}
{Stevenson} K.~B.,  et~al., 2017, \mn@doi [\aj] {10.3847/1538-3881/153/2/68},
  \href {http://adsabs.harvard.edu/abs/2017AJ....153...68S} {153, 68}

\bibitem[\protect\citeauthoryear{{Tennyson} et~al.,}{{Tennyson}
  et~al.}{2016}]{tennyson2016}
{Tennyson} J.,  et~al., 2016, \mn@doi [Journal of Molecular Spectroscopy]
  {10.1016/j.jms.2016.05.002}, \href
  {http://adsabs.harvard.edu/abs/2016JMoSp.327...73T} {327, 73}

\bibitem[\protect\citeauthoryear{{Trotta}}{{Trotta}}{2008}]{trotta_2008}
{Trotta} R.,  2008, \mn@doi [Contemporary Physics] {10.1080/00107510802066753},
  \href {http://adsabs.harvard.edu/abs/2008ConPh..49...71T} {49, 71}

\makeatother
\end{thebibliography}








\bsp	
\label{lastpage}
\end{document}